\documentclass[manuscript,authorversion,nonacm]{acmart}

\AtBeginDocument{%
  \providecommand\BibTeX{{%
    \normalfont B\kern-0.5em{\scshape i\kern-0.25em b}\kern-0.8em\TeX}}}

\usepackage{algorithm}
\usepackage{algorithmic}
\usepackage{hyperref}
\usepackage{pgfplots}
\usepackage{xcolor}
\definecolor{darkblue}{rgb}{0.0, 0.0, 0.7}
\definecolor{darkred}{rgb}{0.8, 0.0, 0.0}
\definecolor{grey}{rgb}{0.5, 0.5, 0.5}

\definecolor{grey_rt_network}{HTML}{87898A}
\definecolor{orange_rt_network}{HTML}{d66f38}
\definecolor{blue_rt_network}{HTML}{384ad6}

\definecolor{blue}{HTML}{4744c9}
\definecolor{green}{HTML}{1e8b6d}
\definecolor{orange}{HTML}{b9601b}
\definecolor{lightblue}{HTML}{32a0a8}
\definecolor{lightgrey}{HTML}{949191}

\definecolor{green_rt}{HTML}{68b36d}
\definecolor{red_rt}{HTML}{c8694b}
\definecolor{grey_rt}{HTML}{a8a6a9}

\hypersetup{
    colorlinks=true,
    linkcolor=black,
    filecolor=magenta,      
    urlcolor=blue,
    citecolor=black,
}

\usepackage{natbib}
\usepackage{graphicx}
\usepackage{subcaption}
\usepackage{float}
\usepackage{lipsum}
\usepackage{enumitem}

\setcitestyle{square, numbers}

\begin{document}

\title[Misinformation and Polarization around COVID-19 vaccines in France, Germany, and Italy]{Misinformation and Polarization around COVID-19 vaccines\\in France, Germany, and Italy}

 \author[ ]{Gianluca Nogara}
 \email{gianluca.nogara@supsi.ch}
 \affiliation{
 \institution{University of Applied Sciences and Arts of Southern Switzerland}
 \country{Switzerland}}
 \author[ ]{Francesco Pierri}
 \email{francesco.pierri@polimi.it}
 \affiliation{
 \institution{Politecnico di Milano}
 \country{Italy}}
 \author[ ]{Stefano Cresci}
 \email{stefano.cresci@iit.cnr.it}
 \affiliation{
 \institution{IIT-CNR}
 \country{Italy}}
  \author[ ]{Luca Luceri}
 \email{lluceri@isi.edu}
 \affiliation{
 \institution{USC Information Sciences Institute}
 \country{USA}}
 \author[ ]{Silvia Giordano}
 \email{silvia.giordano@supsi.ch}
 \affiliation{
 \institution{University of Applied Sciences and Arts of Southern Switzerland}
 \country{Switzerland}}
 
\renewcommand{\shortauthors}{Nogara, et al.}

\begin{abstract}
The kick-off of vaccination campaigns in Europe, starting in late December 2020, has been followed by the online spread of controversies and conspiracies surrounding vaccine validity and efficacy. We study Twitter discussions in three major European languages (Italian, German, and French) during the vaccination campaign. Moving beyond content analysis to explore the structural aspects of online discussions, our investigation includes an analysis of polarization and the potential formation of echo chambers, revealing nuanced behavioral and topical differences in user interactions across the analyzed countries. Notably, we identify strong anti- and pro-vaccine factions exhibiting heterogeneous temporal polarization patterns in different countries. Through a detailed examination of news-sharing sources, we uncover the widespread use of other media platforms like Telegram and YouTube for disseminating low-credibility information, indicating a concerning trend of diminishing news credibility over time. 
Our findings on Twitter discussions during the COVID-19 vaccination campaign in major European languages expose nuanced behavioral distinctions, revealing the profound impact of polarization and the emergence of distinct anti-vaccine and pro-vaccine advocates over time.

\noindent\textcolor{red}{Please remember to cite the published version of this paper:}

\textbf{\textcolor{red}{"Gianluca Nogara, Francesco Pierri, Stefano Cresci, Luca Luceri, Silvia Giordano, "Misinformation and Polarization around COVID-19 vaccines in France, Germany, and Italy" Proceedings of the 16th ACM Web Science Conference 2024"}}
\end{abstract}

\keywords{COVID-19, vaccine, disinformation, misinformation, social media, echo-chamber}

\date{}

\maketitle

\section{Introduction}
Digital platforms played a central role in the dissemination of news and information during the COVID-19 pandemic \citep{dias2020social}.
While online social networks such as Twitter (currently X) provide open and immediate forums for discussion, they also expose their users to abusive behavior and harmful content \cite{SocialFreespeechMisinfo,nogara2023toxic}. An infodemic of false and misleading content~\cite{cinelli2020covid} has likely had a negative impact on vaccination campaigns around the world, contributing to an increase in the number of people who are skeptical and in denial about COVID-19 and the related vaccine \cite{loomba2021measuring,santiago2022covid}. 

The viral spread of online misinformation regarding COVID-19 vaccines posed a serious threat to the safety of many people, as digital platforms are often designed to maximize user engagement~\cite{luceri2021social} and often serve as an important vector in the spread of fake news and conspiracy theories \cite{nogara2022disinformation,newslikevirus,vishnuprasad2024tracking}. Social media platforms also contribute to the creation of user groups that are bound together and hostile toward users with opposing views and opinions \cite{wang2023identifying}, leading to increased polarization among users and the creation of large anti-vaccine groups \cite{jiang2021social}. 
During the COVID-19 pandemic, European countries acted together in order to swiftly develop safe and effective vaccines, with the first COVID-19 vaccination programs kicking off in late December 2020, less than a year into the crisis \cite{EUcommonworkCovid}. 

In this work, we explore the landscape evolution of online vaccine discussions in three European languages: French, German, and Italian. 
In particular, we address the following research questions (RQs):
\begin{itemize}
    \item[\textbf{RQ1:}] \emph{Do we observe the presence and formation of echo chambers? Is there an increase in network polarization over time? }
    \item[\textbf{RQ2:}] \emph{What is the prevalence of low-credibility information in online discussions around vaccines? What are the main vectors for the proliferation of unreliable content?}
\end{itemize}
To answer these questions, we draw on a large-scale dataset of over 30M tweets related to COVID-19 vaccines collected over a period of one year. We study behavioral differences and similarities in the vaccine discourse across countries by conducting a study of user interactions in a dynamic (i.e., time-varying) fashion, providing the following contributions:
\begin{itemize}
    \item We studied the formation of groups of anti- and pro-vaccination supporters by conducting a temporal study of the interaction networks, showing that opposing communities tend to become progressively polarised over time.
    \item We analyzed the prevalence of news from low-credibility sources through the use of fact-checking and media reliability services, uncovering a downward trend in information quality in all considered countries over time.
    \item We carried out a cross-media analysis of some of the main social media used to disseminate fake news \cite{herasimenka2023telegram, lie2020youtube}, characterizing their content and contextualizing their use.
\end{itemize}

\section{Related Work}
\subsection{COVID-19 misinformation}
The vaccine campaign against COVID-19 that kicked off in Europe in December 2020 is situated in a historical period marked by the spread of misinformation, defined as `an era of fake news' \cite{wang2019misinformation}. It is, therefore, important to shed light on how users on social media have handled the discussion.
Research shows how a handful of users can flood social media platforms with questionable content and receive massive endorsement \cite{nogara2022disinformation, cheng2021twittervsfacebook}. A case in point is represented by the Disinformation Dozen, responsible for spreading fake news related to COVID-19 and its vaccine around the world \cite{nogara2022disinformation}.
These actions have not only the ultimate effect of manipulating public opinion on social media but can also lead to offline repercussions for people's health, e.g., creating fear and distrust in vaccine administration \cite{pierri2023one}, or decreasing vaccination coverage and increasing the health risk of exposed people~\cite{loomba2021measuring}.

The debate related to vaccination has already shown that it is an argument subject to the formation of opposing user factions. \citet{cossard2020echo} showed how echo chambers are created in Italy in the debate related to measles based on pro-vax and no-vax ideology. Another study looking at Italy \cite{crupi2022echoes}, this time studying the debate related to the anti-COVID-19 vaccine, showed the division there is between supporters and individuals hesitant to continue during the vaccination campaign, yet found a possible common set of facts on which the two sides could agree. Other studies, such as \cite{lenti2023misinformation}, consider multilingual data and show how the COVID-19 pandemic has led the no-vax communities to become more central to country-specific debates, forming a global Twitter network against vaccination.
Political components in discourse have also been highlighted in the debate, \cite{jiang2020political} conducted a study in COVID-19 conversations in the United States, showing that they are largely influenced by political bias.

\subsection{Vectors of Misinformation: Echo Chambers and Multimedia Platforms}
Meanwhile, there is also a growing interest in studying how online user interactions evolve over time, since such evolution is linked to the dynamics of interactions and to the formation of echo chambers~\cite{tardelli2023temporal}. Among the studies on the structure of networks and their communities \cite{memon2020} suggests that anti-vaxxer communities tend to be denser and more organized than pro-vaxxer communities.

Moving beyond a content-based analysis, \cite{malova2021vaccine} studied more than just the content of the data but also investigated the structure of the network and its characteristics, identifying four different and unconnected discussion clusters of potentially influential individuals capable of manipulating public opinion, showing the ease with which they generate content and become influential.
Others focused on temporal analyses, such as \cite{guntuku2021temporal} that analyzed discussions about the COVID-19 vaccine in the US, studying the variation of vaccination topics on Twitter from December 2020 to February 2021, showing not only a significant temporal variation but also a different attitude among different communities.

Other studies have focused on the potential vectors that have been used for sharing false or misleading news. Among them, \cite{lie2020youtube} found that over a quarter of the most viewed YouTube videos on COVID-19 contained misleading information, reaching millions of viewers worldwide and creating downstream risks for viewers who use YouTube as a source of information. \cite{donzelli2018vaccine} suggests that this trend is not only related to the COVID-19 pandemic, but it is a growing phenomenon already noted in the past. Indeed, the study draws attention to the fact that most of the videos were negative in tone and that the annual number of uploaded videos has increased. Another platform known to hold anti-vaxxer content is Telegram, which presents itself as an unmoderated platform \cite{herasimenka2023telegram}. Focusing on users, far-right groups and COVID-19 protest groups emerged \cite{cliona2022telegram}, interlacing far-right discourse in COVID-19 protest groups.

Given these premises, our objective is to conduct an in-depth examination of the phenomenon of misinformation and user biases within a dataset centered on discussions about COVID-19 vaccines in Europe. This study affords us the opportunity to explore the structure and external media employed in online conversations surrounding COVID-19 vaccines on social media.
An initial investigation was carried out on the existing data, involving only descriptive and statistical analyses \cite{giovanni2022vaccineu}, including hashtags, news sources, and geolocation, serving as an initial foundation for further studies akin to the one presented here.

\section{Data}
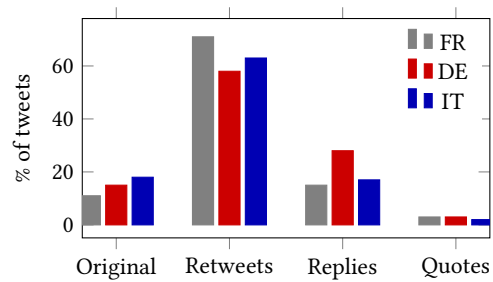
\begin{figure}[t]
\centering
\begin{tikzpicture}
    \begin{axis}[
            ybar,
            symbolic x coords={Original, Retweets, Replies, Quotes},
            xtick=data,
            ylabel={\% of tweets},
            y label style={yshift=-12pt},
            legend style={draw=none},
            bar width = 8pt,
            width=7cm,
            height=4.5cm
        ]
        \addplot[draw=grey, fill=grey] table[x=xaxis,y=FR]{\activitydata};
        \addplot[draw=darkred, fill=darkred] table[x=xaxis,y=DE]{\activitydata};
        \addplot[draw=darkblue, fill=darkblue] table[x=xaxis,y=IT]{\activitydata};
        \legend{FR,DE,IT}
    \end{axis}
\end{tikzpicture}
\caption{Distribution of different kinds of tweets shared for French data (\emph{FR}), German data (\emph{DE}), and Italian data (\emph{IT}).}
\label{ITFRDEactivity}
\end{figure}

We rely on a multilingual dataset of vaccine-related conversations on Twitter, totaling more than 30 million tweets collected using the Twitter Historical API from November 2020 to June 2021 \cite{giovanni2022vaccineu}, a period which covers the start of vaccination campaigns in European countries. The dataset consists of online discussions around the COVID-19 vaccine on Twitter in three major European languages (French, German, and Italian).\footnote{For more details on the data collection, we refer the reader to the related paper \cite{giovanni2022vaccineu}.} For the sake of simplicity, in our analyses, we will refer to the three European countries in which these languages are spoken the most, namely France, Germany, and Italy, although the data might capture conversations from other countries as well (e.g., Austria, Canada, Switzerland, etc.). Figure \ref{ITFRDEactivity} shows the distribution of different tweets shared by users in our dataset, with the large majority composed of retweets in all three languages. 

\pgfplotstableread[row sep=\\,col sep=&]{
    xaxis & FR & DE & IT \\
    Original & 11 & 15 & 18\\
    Retweets & 71 & 58 & 63\\
    Replies & 15 & 28 & 17\\
    Quotes & 3 & 3 & 2\\
    }\activitydata

\begin{figure}[h!]
  \centering
  \begin{subfigure}{0.45\textwidth}
    \includegraphics[width=\linewidth]{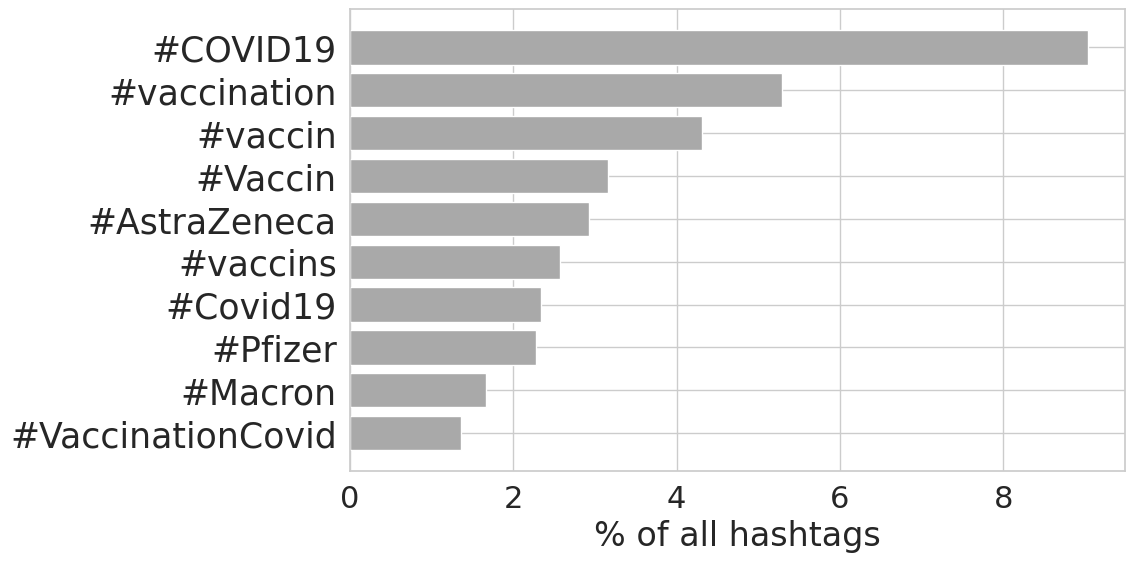}
    \caption{French dataset}
  \end{subfigure}
  \begin{subfigure}{0.45\textwidth}
    \includegraphics[width=\linewidth]{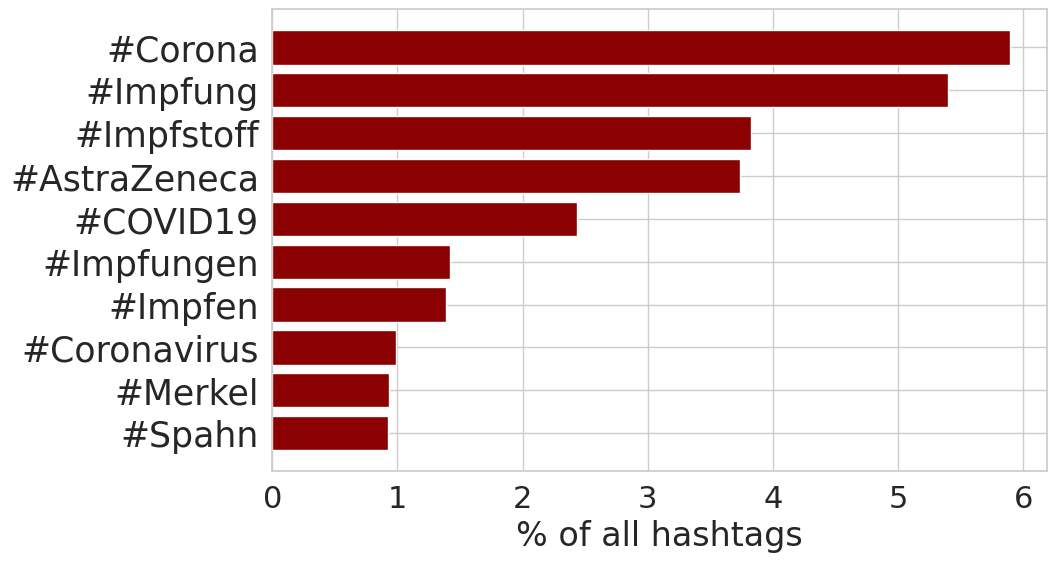}
    \caption{German dataset}
  \end{subfigure}
  \begin{subfigure}{0.45\textwidth}\textbf{}
    \includegraphics[width=\linewidth]{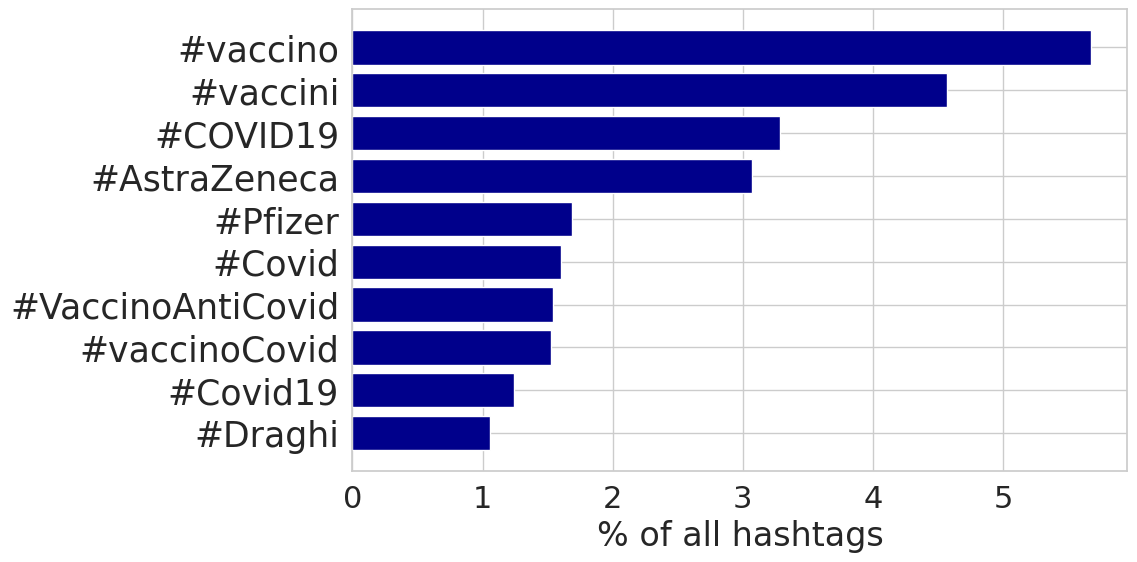}
    \caption{Italian dataset}
  \end{subfigure}
  \caption{Top-10 most frequent hashtags by country.}
  \label{hashtags}
\end{figure}

As an exploratory step, we look at the prevalence of different hashtags, which provides insights into the discussions that have engaged users in each country. Figure \ref{hashtags} shows the top 10 most used hashtags, highlighting how the main topics of discussion were those related to vaccines and COVID-19. However, for all three datasets, we also find hashtags related to the political sphere. For example, we have \emph{\#Macron} for the French language dataset, \emph{\#Draghi} for the Italian language dataset, \emph{\#Merkel} and \emph{\#Spahn} for the German language dataset, highlighting how despite the differences between countries we find rather common behavioral patterns. Also of importance is \emph{\#AstraZeneca}, which stands out as a highly used hashtag in all three datasets, remarking how much impact AstraZeneca's discontinuation of the anti-COVID-19 vaccine has had in public opinion \cite{JEMIELNIAK20214}.

\begin{figure}[t]
     \centering
     \includegraphics[width=\linewidth]{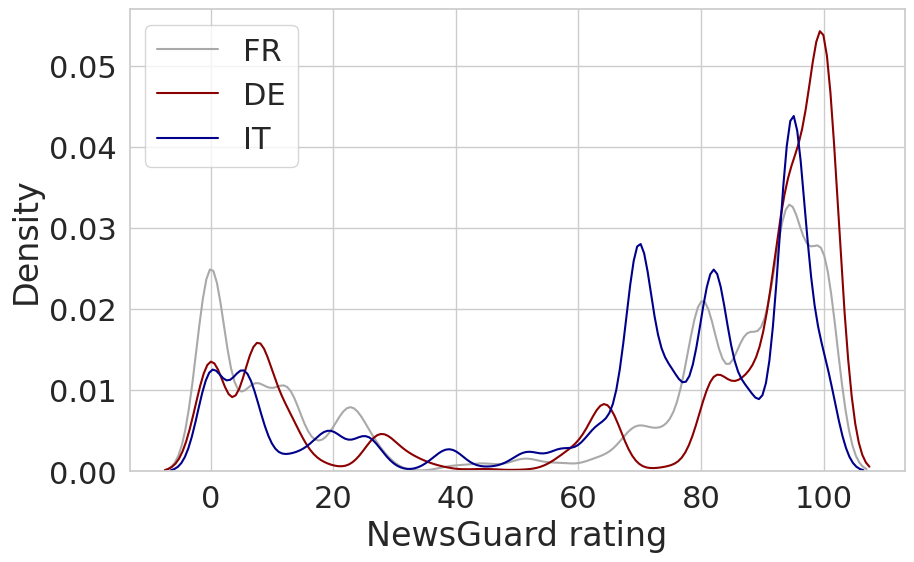}
     \caption{Density distribution of reliability ratings of tweets containing a domain classified by NewsGuard, estimated by Kernel Density Estimation (KDE), for each country.}\label{density_ratings}
\end{figure}

\section{Methodology}
\subsection{Reliability of Information Sources}
To ensure a comprehensive and reliable assessment of the credibility of news outlets shared along vaccine-related content, we used a distant-supervision approach to label websites adopted in the literature \cite{pierri2020diffusion,cheng2021twittervsfacebook,pierri2023one}. For the labels, we employed NewsGuard\footnote{https://www.newsguardtech.com/} ratings, given its established reputation as an independent and transparent organization that employs human experts in the field of journalism. NewsGuard conducts rigorous evaluations of news sources, considering factors such as transparency, accountability, adherence to journalistic standards and error correction, and promoting high-quality information \cite{reuters2019mediaguard}. We used the service via an API that provides credibility ratings for news outlets, with scores ranging from 0 to 100, where 0 indicates a completely unreliable news outlet, and 100 is a completely trustworthy one. Specifically, we extracted all URLs from the tweets in our dataset and we queried the API to obtain credibility scores for more than 50\% of all URLs mentioned in our datasets, retrieving a score for 
    65\% of all URLs in the French dataset, 70\% of all URLs in the German dataset, 
    and 57\% of all URLs in the Italian dataset.
Since NewsGuard provides various reliability ranges \cite{newsguardprocesscriteria}, we classify domains into two categories: high credibility and low credibility. To select a threshold to determine whether or not a news outlet is credible, we analyzed the distributions of ratings across our datasets.
As shown in Figure \ref{density_ratings}, the distributions of NewsGuard credibility ratings for the three datasets tend to be bimodal. 
In fact, we find a large number of domains whose credibility ratings are $\ge60$ and $\le30$. 
Following \cite{pierri2023one}, we used a conservative criterion and labeled those news outlets with NewsGuard rating $\leq$ 30 as \textit{low credibility}, and those with rating $\geq$ 60 as \textit{high credibility}.
\begin{figure*}[h!]
  \centering
  \begin{subfigure}{0.290\textwidth}
    \includegraphics[width=\linewidth]{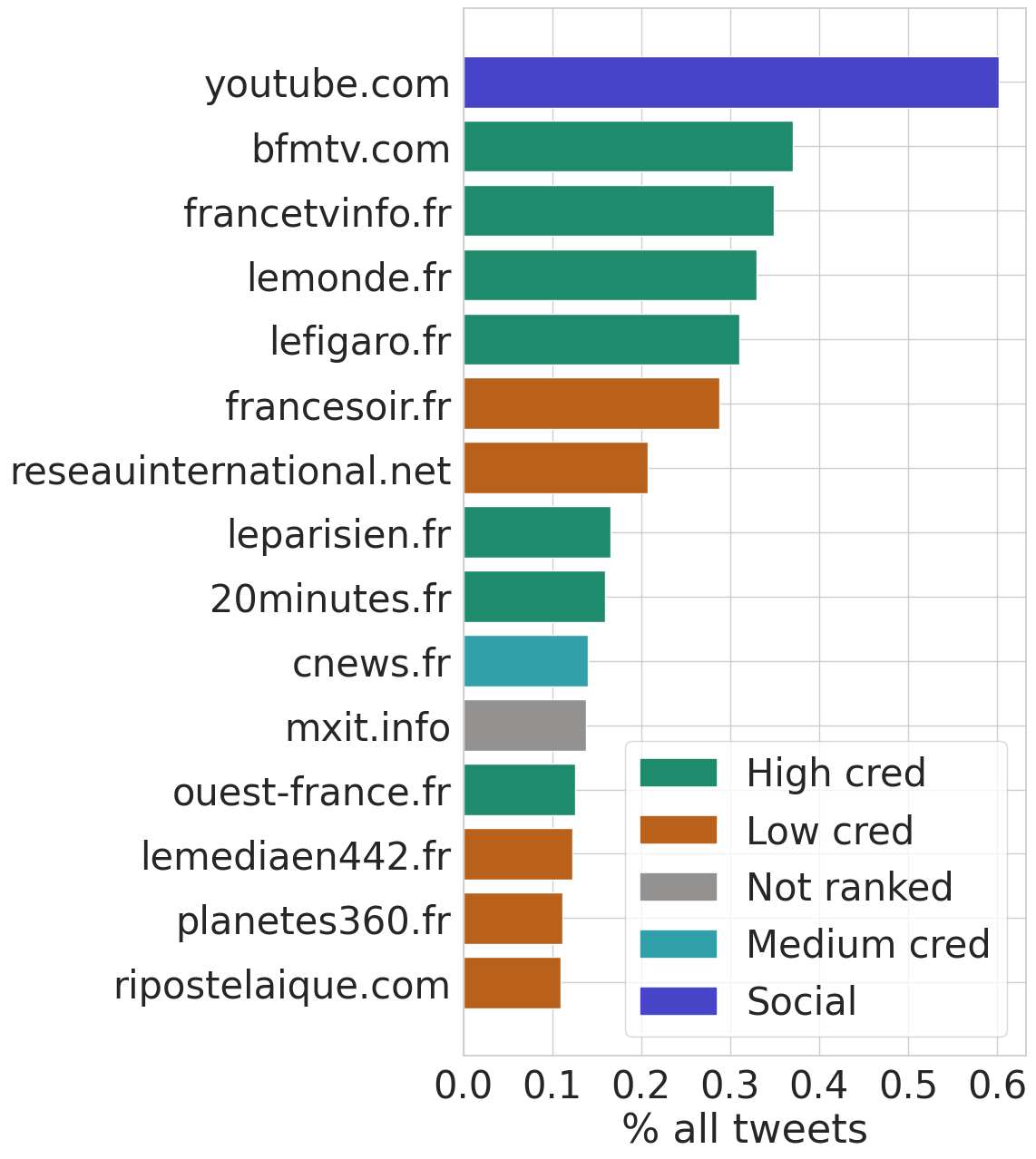}
    \caption{French dataset}\label{fr_domain}
  \end{subfigure}
  \hspace{0.05\textwidth}
  \begin{subfigure}{0.25\textwidth}
    \includegraphics[width=\linewidth]{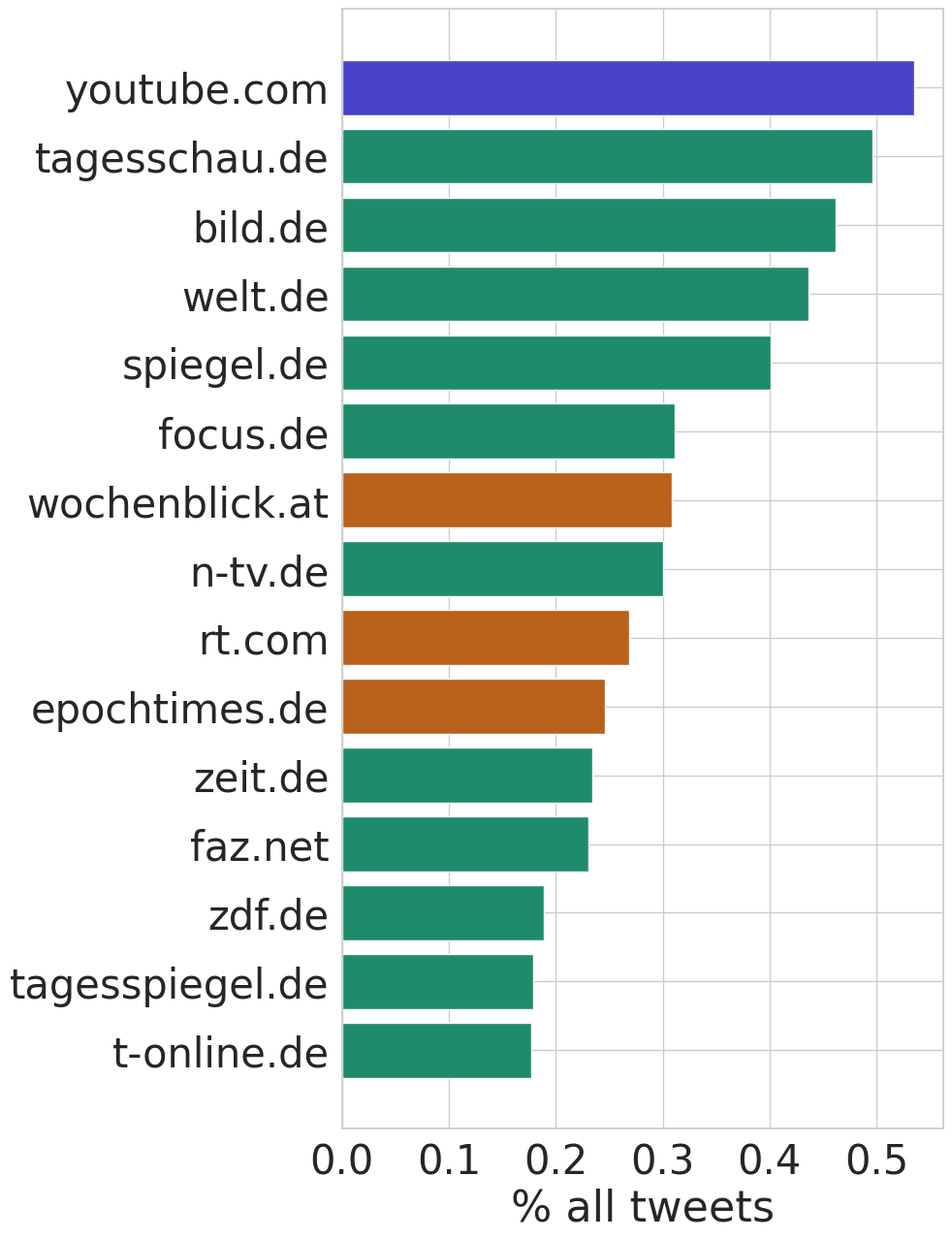}
    \caption{German dataset}\label{de_domain}
  \end{subfigure}
  \hspace{0.05\textwidth}
  \begin{subfigure}{0.265\textwidth}
    \includegraphics[width=\linewidth]{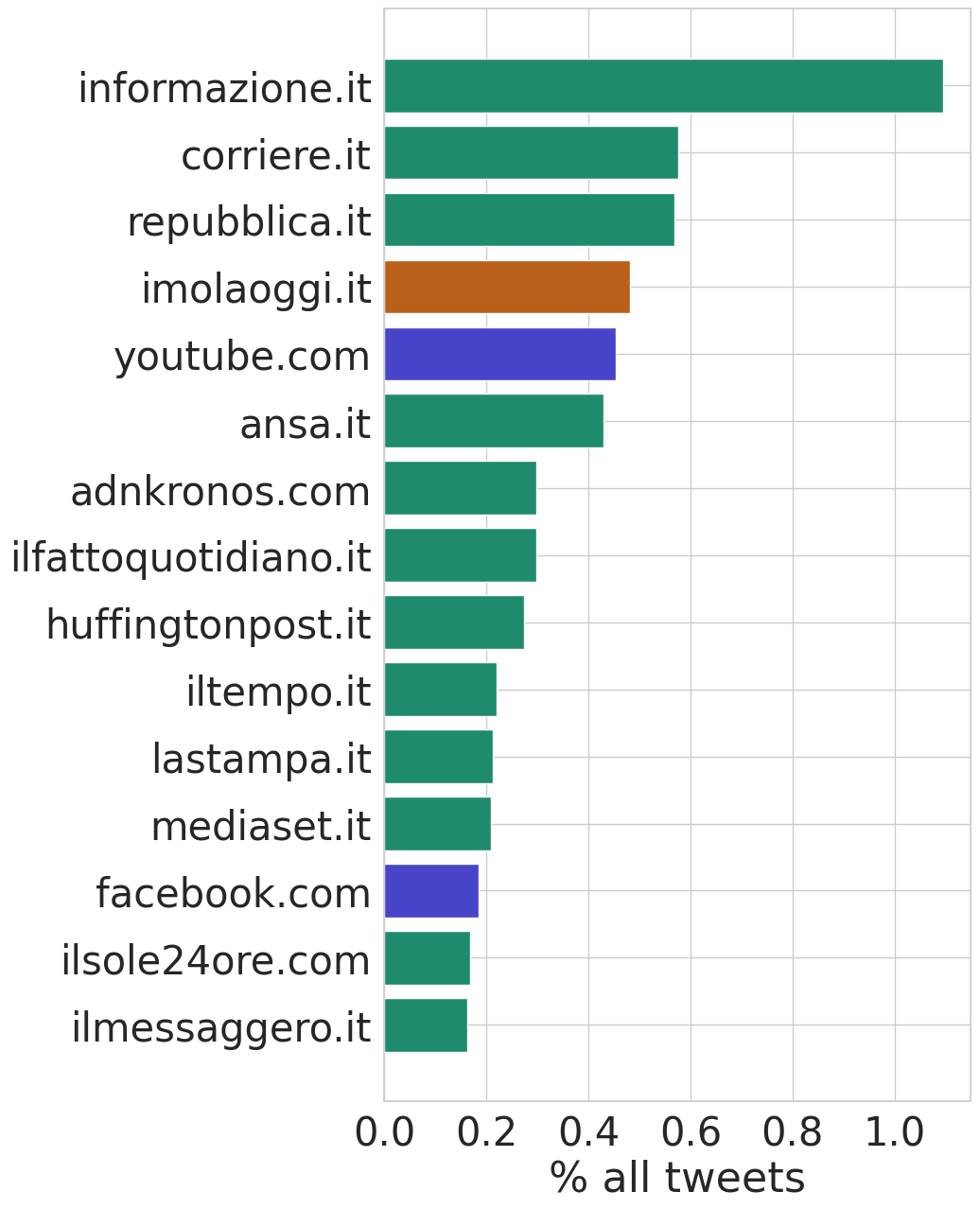}
    \caption{Italian dataset}\label{it_domain}
  \end{subfigure}
  \caption{Top-15 shared domains in each dataset colored according to their NewsGuard rating.}\label{domains}
\end{figure*}

An exploratory analysis of most shared web domains is provided in Figure~\ref{domains}. We observe a large prevalence of YouTube, which is particularly relevant given the role of this platform for spreading fake news and conspiracy theories~\cite{gatta2023interconnected,hussein2020measuring,ginossar2022cross}. We also notice the presence of at least one low-credibility web domain in the top-15 ranking of all three countries.

Given the importance of YouTube links in our datasets, which account for 2.2\% (171.256) of the links shared, we exploited the YouTube Data API to collect information on linked videos. We were able to obtain a range of information such as the availability of a video, as well as its duration, title, description, number of views, number of likes, number of comments, channel, publication date, and the possible presence of a label related to COVID-19~\cite{knuutila2020covid,krishnan2021research}. 

\subsection{Building Retweet Networks}
To analyze the structure of interactions between users in our datasets, we built and analyzed a retweet network for every country.
As retweets often indicate a form of social endorsement \cite{retweetsindicate, metaxas2014retweets}, they represent an effective method to identify groups of users interacting with and approving each other~\cite{tardelli2023temporal}. To study how these users are connected and grouped within the retweet network, we apply the well-known Louvain community discovery algorithm \cite{blondel2008fast}. Since we aim to study how the interactions evolve over time, we built and compared three different snapshots of the interaction networks, each taken at a different point in time. 
These times were chosen to capture peaks of activity corresponding to important real-world events. We consider a 7-day lag from the event for gathering the bulk of tweet cascades about the events, which we use to build our retweet networks. 

\begin{figure}[]
     \centering
     \includegraphics[width=\linewidth]{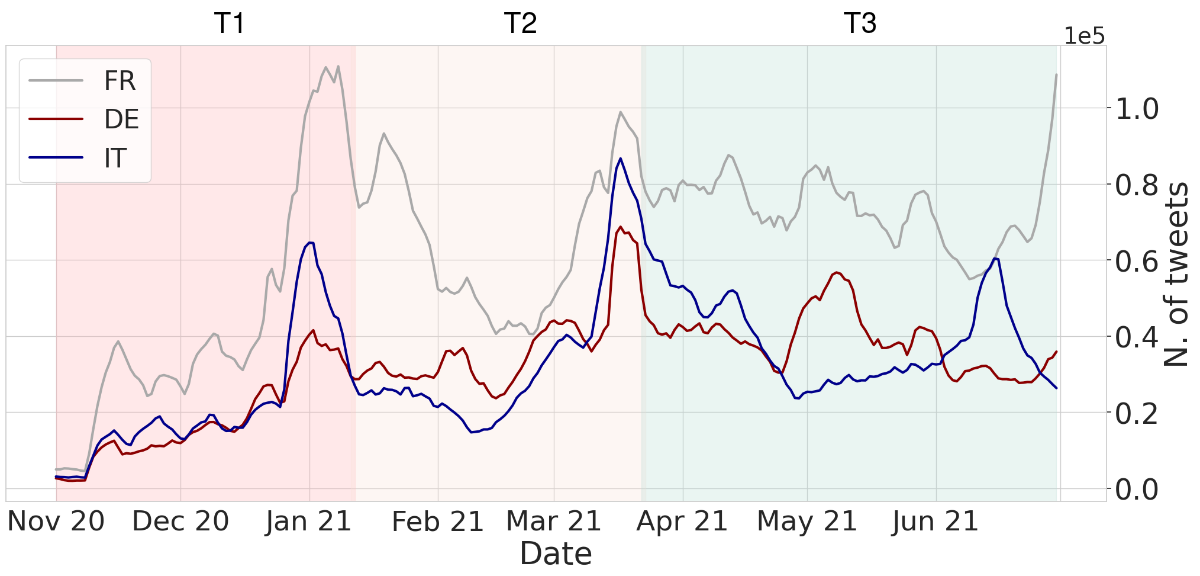}
     \caption{Moving average of the daily number of tweets shared in each dataset. We highlight periods in which we constructed three different retweeting networks with different colors.}\label{snapshot}
\end{figure}

Figure~\ref{snapshot} shows the selected intervals with respect to Twitter activity in the different datasets. We can see similar temporal patterns across languages, with peaks corresponding to significant events such as the start of COVID-19 vaccine administration in Europe on December 27, 2020, commonly referred to as the ``Vaccine Day'', and the suspension of the AstraZeneca vaccine by several European countries (e.g., Italy, France, Germany, Spain, Portugal and Austria) on March 15, 2021. As these events are not only national but also European in character, the patterns in the number of activities across the three datasets are quite consistent. There is an increase in the number of interactions during international events, and minor variations in activity levels can be attributed to secondary or internal factors. 

We considered the following intervals to build three different data snapshots, for each language:
\begin{itemize}
    \item[\textbf{T1}:] [01/11/2020 - 11/01/2021] a week after the peak of activity on January 5th, which corresponds to the beginning of the vaccination campaign;
    \item[\textbf{T2}:]  [12/01/2021 - 31/03/2021] a week after the peak of activity on March 15th, which corresponds to the suspension of AstraZeneca in many European countries, including France, Germany and Italy;
    \item[\textbf{T3}:] [01/04/2021 - 30/06/2021] end of the collection period.
\end{itemize}

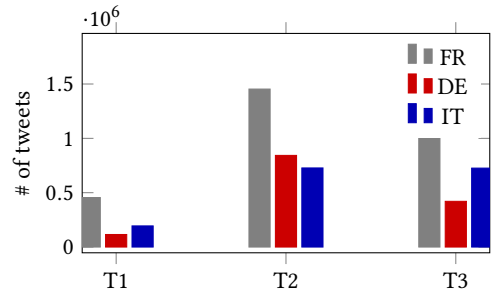
\begin{figure}[t]
\centering
\begin{tikzpicture}
    \begin{axis}[
            ybar,
            symbolic x coords={T1, T2, T3},
            xtick=data,
            ylabel={\# of tweets},
            y label style={yshift=-12pt},
            legend style={draw=none},
            bar width = 8pt,
            width=7cm,
            height=4.5cm
        ]
        \addplot[draw=grey, fill=grey] table[x=xaxis,y=FR]{\activitydata};
        \addplot[draw=darkred, fill=darkred] table[x=xaxis,y=DE]{\activitydata};
        \addplot[draw=darkblue, fill=darkblue] table[x=xaxis,y=IT]{\activitydata};
        \legend{FR,DE,IT}
    \end{axis}
\end{tikzpicture}
\caption{Number of tweets captured in each snapshot for French (\emph{FR}), German (\emph{DE}), and Italian datasets (\emph{IT}).}
\label{ITFRDEtweets}
\end{figure}

\begin{figure}[t]
\centering
\begin{tikzpicture}
    \begin{axis}[
            ybar,
            symbolic x coords={T1, T2, T3},
            xtick=data,
            ylabel={\# of users},
            y label style={yshift=-12pt},
            legend style={draw=none},
            bar width = 8pt,
            width=7cm,
            height=4.5cm
        ]
        \addplot[draw=grey, fill=grey] table[x=xaxis,y=FR]{\activitydata};
        \addplot[draw=darkred, fill=darkred] table[x=xaxis,y=DE]{\activitydata};
        \addplot[draw=darkblue, fill=darkblue] table[x=xaxis,y=IT]{\activitydata};
        \legend{FR,DE,IT}
    \end{axis}
\end{tikzpicture}
\caption{Number of users captured in each snapshot for French (\emph{FR}), German (\emph{DE}), and Italian datasets (\emph{IT}).}
\label{ITFRDEusers}
\end{figure}
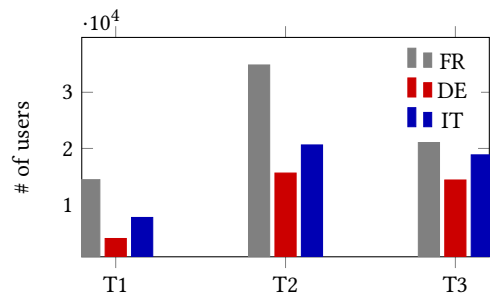

Each snapshot contains a substantial amount of user data and tweets in order to construct and analyze retweet networks. Figures \ref{ITFRDEtweets} and \ref{ITFRDEusers} show the numbers of tweets and users underlying each snapshot. Networks corresponding to different periods might be overlapping in terms of nodes, in case users are active during multiple periods.

\section{Results}


\subsection{RQ1: Polarization}
By building a retweet network for each dataset at different time intervals T1, T2, and T3, we study the evolution of user interactions and the possible formation of communities, or the confluence of smaller communities in those of greater relevance~\cite{tardelli2023temporal,nogara2022disinformation}. 
We analyze and classify clusters of users based on their shared content and interactions. By manually examining the hashtags and texts used by these users, we categorize them within specific clusters. Our study is not solely focused on nodes with the highest centrality or popularity; rather, we randomly select nodes to investigate and categorize their shared content. Analyzing the content shared by the different network communities of the Italian language dataset, in fact, we find that the anti-vaxxer community contains very few political figures. This finding relates to the Italian government's situation during most Italian vaccination campaigns. In fact, the election of a new prime minister (i.e., Mario Draghi), who remained in office from February 13, 2021 to October 22, 2022 led to a government of national unity without real opposition. The situation of the French dataset is significantly different given the French political situation during the period under analysis. The government led by Emmanuel Macron received strong criticism, especially from far-right and left-wing parties, such as \emph{Les Patriotes}, whose leader belongs to the anti-vaxxer community in the French network.

\begin{figure*}[t!]
         \includegraphics[width=0.8\textwidth]{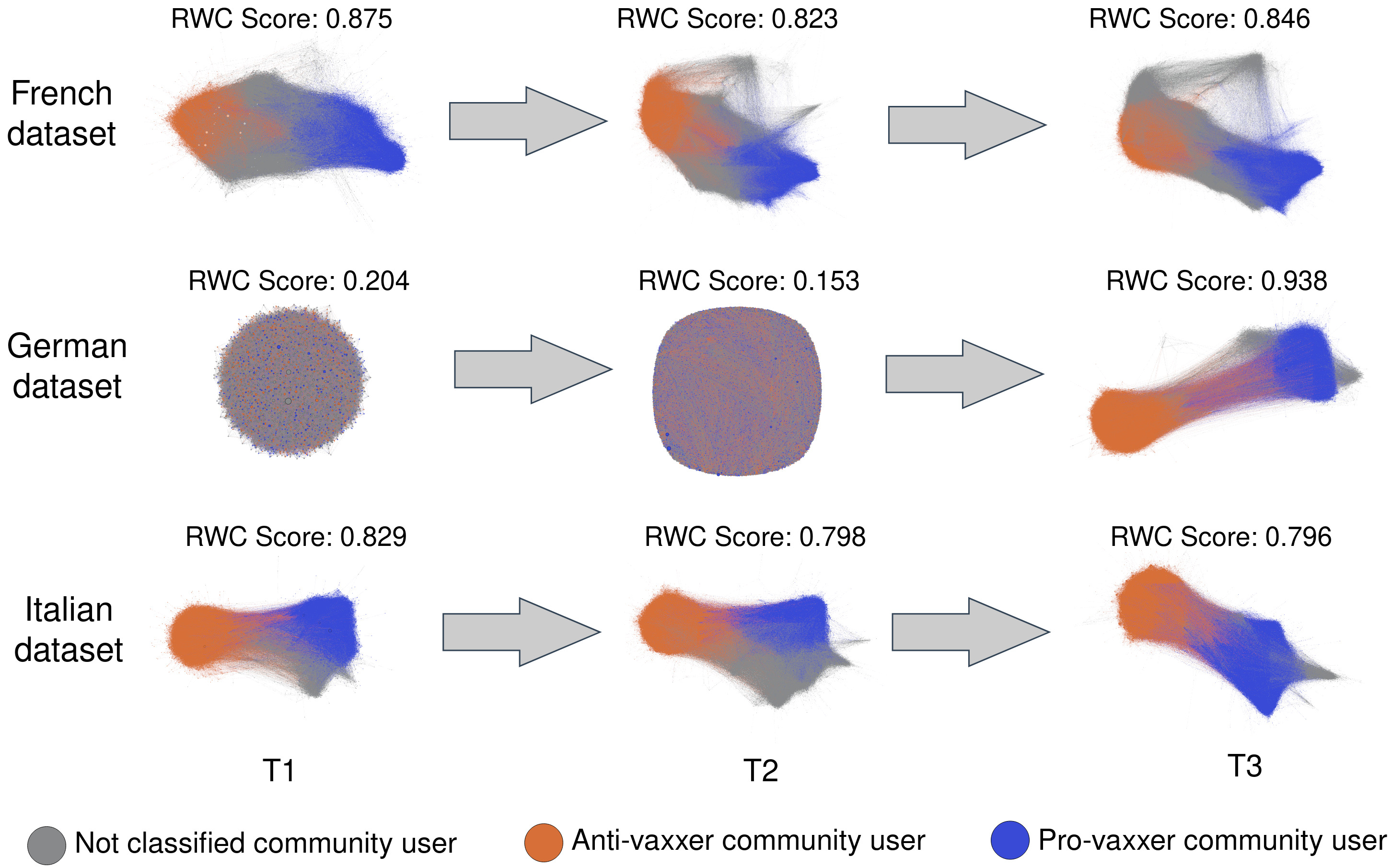}
     \caption{Temporal evolution of retweet networks at different time intervals with related controversy scores (Random Walk Controversy -- RWC) for each country.}\label{evolution}
\end{figure*}

In addition to analyzing the content of the communities, we also applied metrics to define the degree of fragmentation and overall polarization of the networks at each time interval. In literature, Random Walk Controversy (RWC) is a well-known measure that assesses the degree of fragmentation and controversy within polarized discussion networks \cite{garimella2018randomwalk}. We then computed RWC scores at each snapshot and for each network, as reported in our annotations to Figure~\ref{evolution}.

\pgfplotstableread[row sep=\\,col sep=&]{
    xaxis & FR & DE & IT \\
    T1 & 14434 & 3961 & 7707\\
    T2 & 34862 & 15615 & 20622\\
    T3 & 36512 & 14380 & 18855\\
    }\activitydata

\pgfplotstableread[row sep=\\,col sep=&]{
    xaxis & FR & DE & IT \\
    T1 & 455299 & 115348 & 194375\\
    T2 & 1453398 & 843130 & 727668\\
    T3 & 1797118 & 420530 & 726149\\
    }\activitydata

In Figure~\ref{evolution}, we highlight the evolution of the retweet networks at times T1, T2, and T3. At each time step, each node is colored based on the community it belongs to, according to the following scheme:
\begin{itemize}
    \item[\textcolor{blue_rt_network}{\rule{0.25cm}{0.25cm}}] the blue color represents the community characterized by a strong \textit{pro-vaxxer} component;
    \item[\textcolor{orange_rt_network}{\rule{0.25cm}{0.25cm}}] the orange color represents the community characterized by a strong \textit{anti-vaxxer} component;
    \item[\textcolor{grey_rt_network}{\rule{0.25cm}{0.25cm}}] the grey color represents \textit{other} communities.
\end{itemize}

Through a qualitative analysis of user-shared content and user clusters, we were able to characterize the datasets. By performing a manual study of the textual content shared by the most active users and random users within the network. We then labeled the users and clusters according to their feeds, content, and the users that comprise them, so we were able to identify the peculiar dynamics of online debates about vaccination as they unfolded over time in the three countries, which can be summarized as follows:
\begin{itemize}
  \item The French language dataset is characterized by several political communities, with pro-vaxxer users coinciding with the ruling parties and media, and anti-vaxxer users with government opposition groups. This dataset is thus characterized by a strong political component.
  \item The German language dataset is very homogeneous and generally does not have well-defined user communities, which become evident during the last period of analysis (T3).
  \item The Italian language dataset maintains the same structure from the first to the last snapshot, with two large communities of anti-vaxxer and pro-vaxxer users mainly clashing for ideological reasons rather than for political ones.
\end{itemize}

\pgfplotstableread[row sep=\\,col sep=&]{
    xaxis & FR & DE & IT \\
    T1 & 14434 & 3961 & 7707\\
    T2 & 34862 & 15615 & 20622\\
    T3 & 36512 & 14380 & 18855\\
    }\activitydata

Except for the German network, anti-vaxxer and pro-vaxxer communities appear early in the discussion, as shown by their presence at T1 for Italian and French. However, the three networks exhibit a similar structure at T3, when about 80\% of all users either belong to the anti-vaxxer or pro-vaxxer communities.
This kind of behavior leads communities to close outward and generate echo chambers in which tweets are shared by users with similar interests, values, or identities who then tend to aggregate~\cite{villa2021echo,jiang2021social}.

We can observe that the RWC scores for French and Italian exhibit similar high values throughout the period of analysis ($\sim$ 0.8). The  RWC score for German increases from 0.204 to 0.938, showing the formation of two echo chambers, one anti-vaxxer and one pro-vaxxer, which are only evident in the last period of analysis (T3).

\subsection{RQ2: Media Credibility}
Links to external websites are widely shared by users in our dataset, with more than 6,500,000 tweets containing at least one URL (excluding retweets that link to "twitter.com" or "t.co"). We analyze these URLs to better understand the type of content that is shared by users, both in terms of the type of link shared (e.g., to a website or another social network) and the reliability of the external website.

\subsubsection{NewsGuard and source reliability}
\begin{figure}[t]
     \centering
     \includegraphics[width=\linewidth]{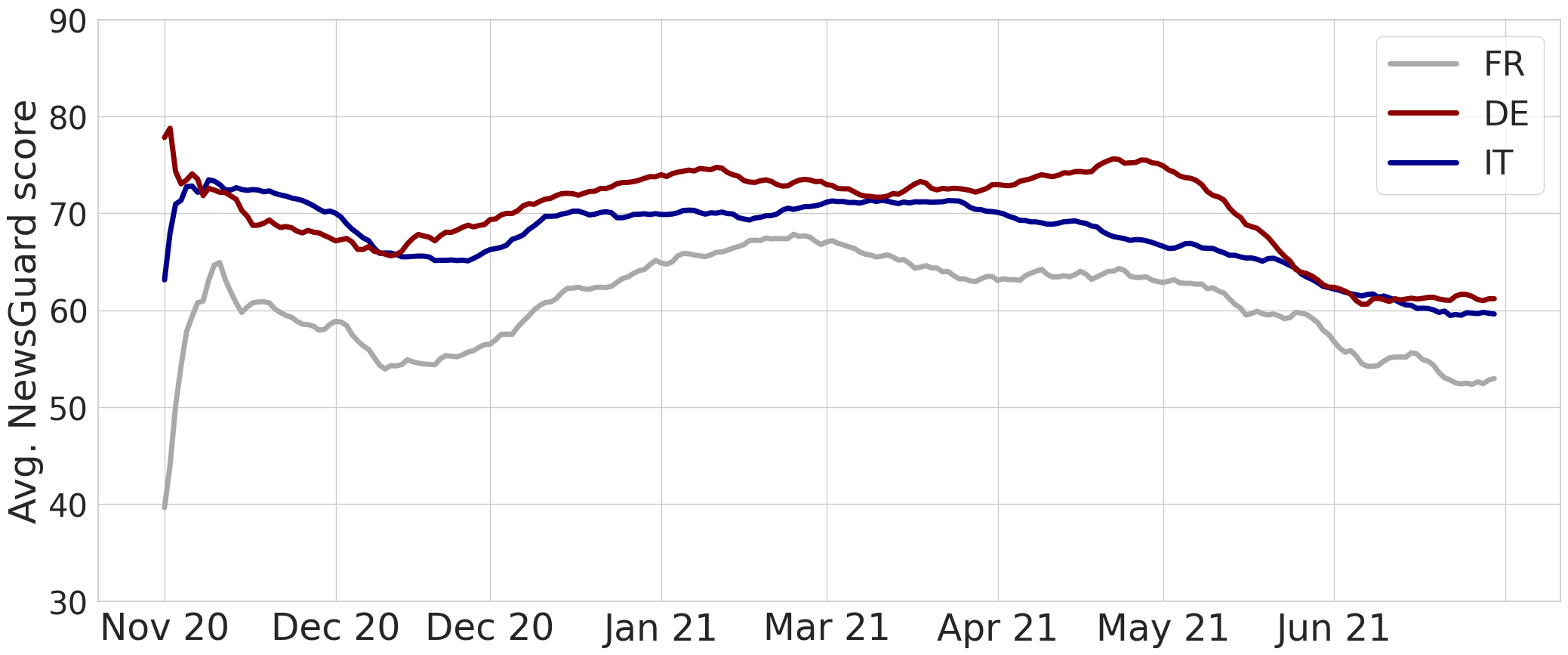}
     \caption{Average daily rating of links to websites on NewsGuard's list.}\label{fig:newsguard}
\end{figure}

We first analyze the average reliability score of news websites shared by Twitter users over time. As shown in Figure \ref{fig:newsguard}, we observe that in the three datasets the average daily consumption of news leans toward reputable sources (FR = 61.0, DE = 70.5, IT = 67.2). We remark that according to NewsGuard, a reputable source has a score higher than 60. We can notice that the reliability of news websites in French conversations about vaccines is smaller compared to Italian and German ones.

\begin{figure}
     \centering
     \includegraphics[width=\linewidth]{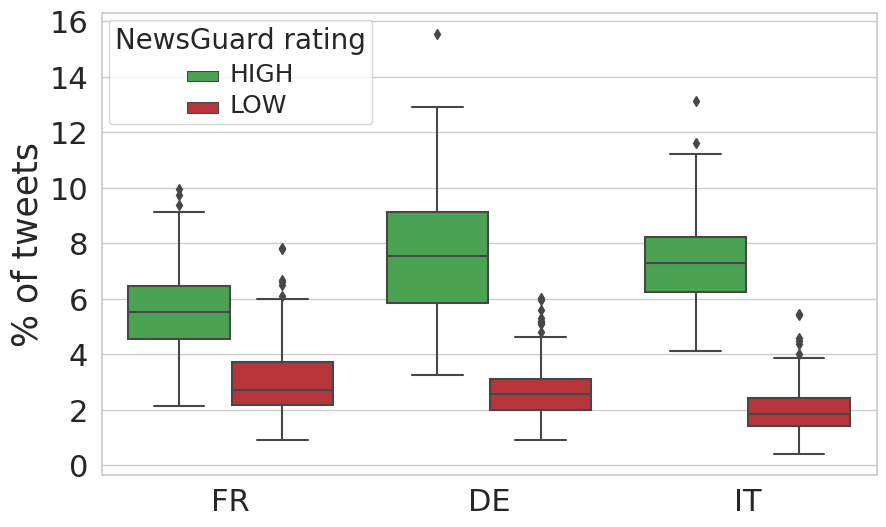}
     \caption{Distributions of the daily proportion of tweets linking to High- and Low-credibility news websites for each dataset. Median HIGH NewsGuard rating \emph{FR} 5.5, \emph{DE} 7.5 and \emph{IT} 7.3, median LOW NewsGuard rating \emph{FR} 2.7, \emph{DE} 2.6 and \emph{IT} 1.8}\label{fig:newsguard_boxplot}
\end{figure}

Next, we analyze the daily prevalence of tweets sharing links to either High- or Low-credibility news websites in the three datasets. Figure \ref{fig:newsguard_boxplot} shows that tweets sharing news articles are a minority (less than 10\% of daily tweets on average), but reputable sources are significantly more prevalent than unreliable ones in all datasets (cf. median values in the caption). In line with previous findings, we notice that low-credibility information makes up a small but not negligible proportion of online sources \cite{pierri2023one}.

We then focus on the reliability of news articles shared by anti-vaxxer and pro-vaxxer communities. Anti-vaxxer communities are often responsible for spreading disinformation and misleading news~\cite{germani2021anti} about vaccines. 

\begin{figure*}
     \centering
     \includegraphics[width=\linewidth]{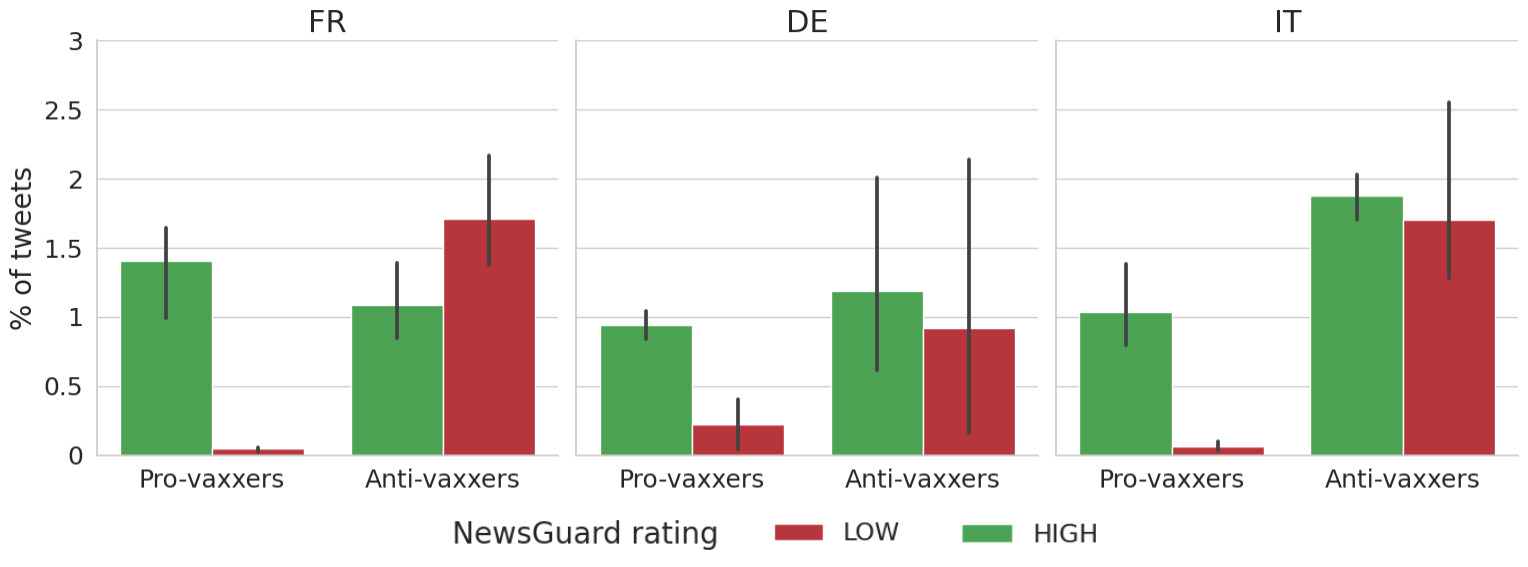}
     \caption{Percentage of tweets linking to High- and Low-credibility news websites for the \emph{anti-vaxxers} and \emph{pro-vaxxers} communities obtained from the retweet network for each dataset. We consider data from all iterations, and the error bar is computed with the discrepancy in each time period T}\label{fig:newsguard_iterations}
\end{figure*}

As we can see from Figure \ref{fig:newsguard_iterations}, pro-vaxxers groups, in all countries, mainly share links referring to high-credibility sources. The links referring to low-credibility sources, both in the French and in the Italian language dataset, is a very minimal percentage, close to zero. And this is valid for all time periods. Only in the German dataset we observe an increase in the error bar due to the more evolving nature of this dataset.
The situation in anti-vaxxers communities see a very strong number of links referring to low-credibility sources. Interestingly, the anti-vaxxers communities identified in the French language dataset show a prevalence of links referring to low-credibility sources, suggesting a strong presence of disinformation. In all three dataset, we see a high variability along time (large error bar) as we can observe an increasing polarization in the anti-vax discussion in all languages.


\subsubsection{Links to Other Media Platforms}
We investigate whether YouTube plays a role in promoting false and misleading claims about vaccines, as often reported in literature \cite{lie2020youtube,hussein2020measuring,nogara2022disinformation}, by extracting metadata about videos shared by Twitter users in the three datasets.

\begin{figure}[t]
     \centering
     \includegraphics[width=\linewidth]{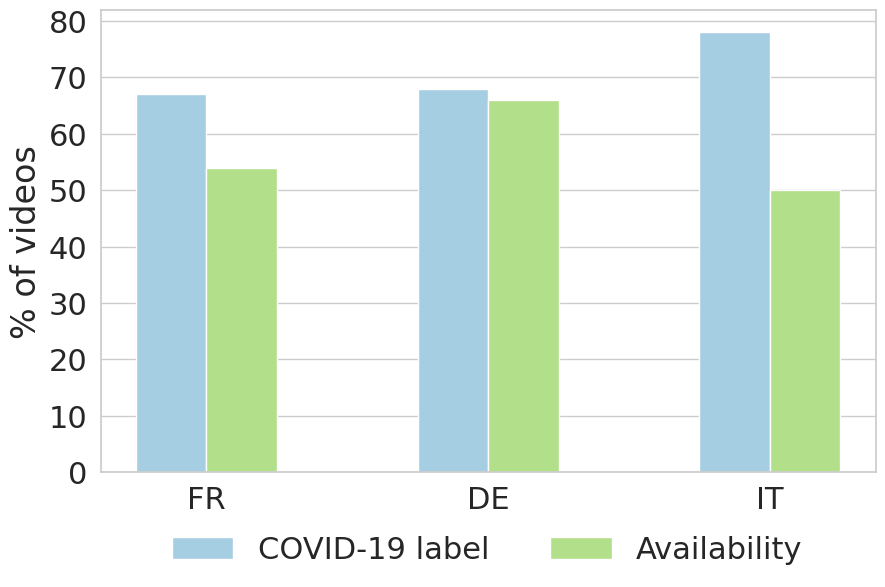}
     \caption{Proportion of YouTube videos shared on Twitter that had a COVID-19 label (blue) and those that are still available (green) as of May 2023, for each dataset.}\label{youtube_stats}
\end{figure}

\begin{table}[!t]
\caption{Stats related to the NewsGuard rating of domains of users who share YouTube links .}
\label{newsguard_youtube}
\begin{tabular}{c|c|c}
\hline
\textbf{Dataset} & \textbf{Rating (all users)} & \textbf{Rating (users sharing YT)}\\ \hline
FR               & 62.0                                & 46.4\\
DE               & 70.5                                & 56.1\\
IT               & 67.2                                & 53.6\\
\hline
\end{tabular}
\end{table}

We first look at the prevalence of COVID-19 labels attached to videos by the platform as well as their availability at the time of the analysis (May 2023). 
COVID-19 labels has been placed on all videos that have COVID-19 as a topic of discussion in 2021. This label leads back to the WHO website. The purpose of this action is to counter misinformation on the platform \cite{elias2021youtube}.
The unavailability of a video refers to a suspension of the channel, a suspension of the video, or the author deleting the video from the platform. Unavailable videos have been linked to sources of misinformation in the past \cite{pierri2023one,cheng2021twittervsfacebook}. As shown in
Figure \ref{youtube_stats}, over 60\% are related to COVID-19 vaccines, as expected, given the topic of conversations. Approximately 50\% of the videos in French and Italian are no longer available, while almost 70\% are still available in German. This suggests a higher prevalence of unreliable content originating on YouTube in French and Italian than in German.

\begin{figure}[t]
     \centering
     \includegraphics[width=\linewidth]{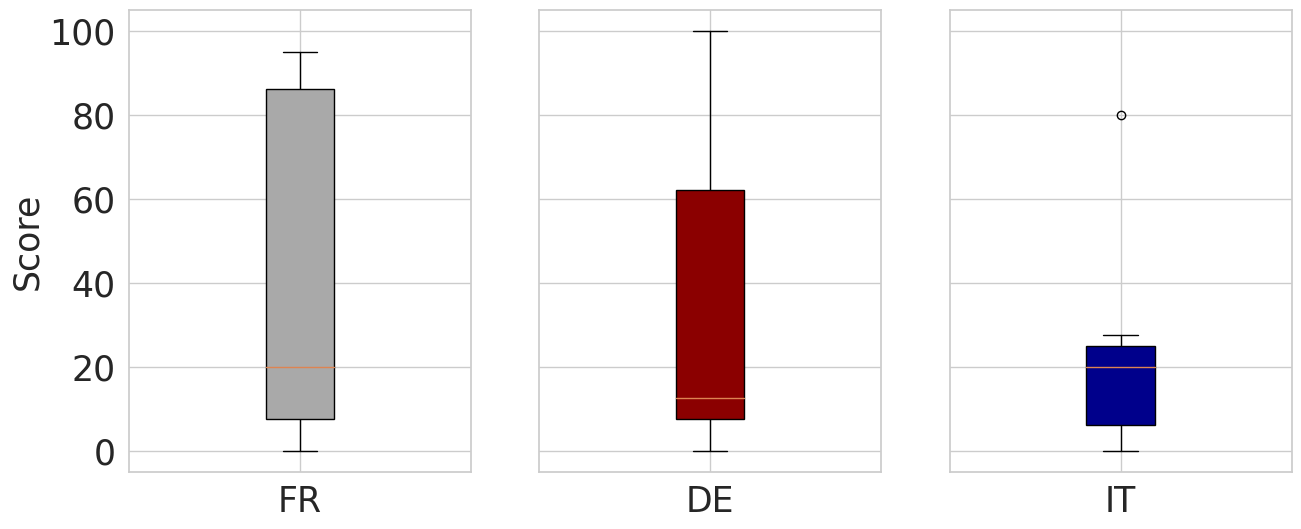}
     \caption{Distributions of reliability ratings (NewsGuard) of domains closely related to YouTube, as derived from the weighted network analysis, for each dataset.}\label{fig:boxplot}
\end{figure}

Through a qualitative study, we identified several YouTube channels related to anti-vax movements among those most shared by users in all countries. To achieve this, all links related to YouTube videos were filtered from the data by going to the most re-shared videos and studying the channels that posted them in order to get a clearer view of the content and type of channels. We then investigate the reliability of news sources shared by users who shared YouTube videos, finding that such users tend to share less reliable news articles, as shown in Table \ref{newsguard_youtube}. To corroborate this analysis, we perform a co-occurrence analysis on the bipartite network of users and their shared URLs. A projection of this bipartite graph produces a co-occurrence weighted network whose weight between URLs $i$ and $j$ represents the total number of users that shared both $i$ and $j$. Next, we assign the NewsGuard score of the website to each URL, if available. For all YouTube URLs, we can thus extract a reliability score by taking the weighted average score of all its neighbors. Figure \ref{fig:boxplot} shows the distribution of such YouTube reliability scores for all datasets. We can notice that the median values are very low in all cases (FR = 20, DE = 12.5, and IT= 20.0), further indicating that YouTube URLs are often shared by users actively engaging with unreliable news sources \cite{gatta2023interconnected}.


A similar analysis was carried out on the shared Telegram channels, which might be very relevant vectors of misinformation since they are not moderated in any way~\cite{ng2020analyzing, herasimenka2023telegram}. 
Telegram links, which account for 0.16\% (12.395) of the links shared, allowed us to get the information and content shared by different public channels. We obtained 492 channels and 4,714,193 messages (62 channels with 628,548 messages in the French dataset, 324 channels with 2,607,947 messages in the German dataset, and 106 channels with 1,477,698 messages in the Italian dataset).
From the point of view of shared messages, the channels in the German-language dataset have the highest volume; however, the channels in the Italian-language dataset have the highest ratio of message per channel. A manual qualitative inspection indicates that most shared channels from all datasets are mainly from anti-vaxxer or conspiracy groups, involving more than one million users. 
Through NewsGuard, we obtain the rating of domains shared within Telegram channels, similar to the rating of links shared by the users in the Twitter datasets.


\begin{figure}
     \centering
     \includegraphics[width=\linewidth]{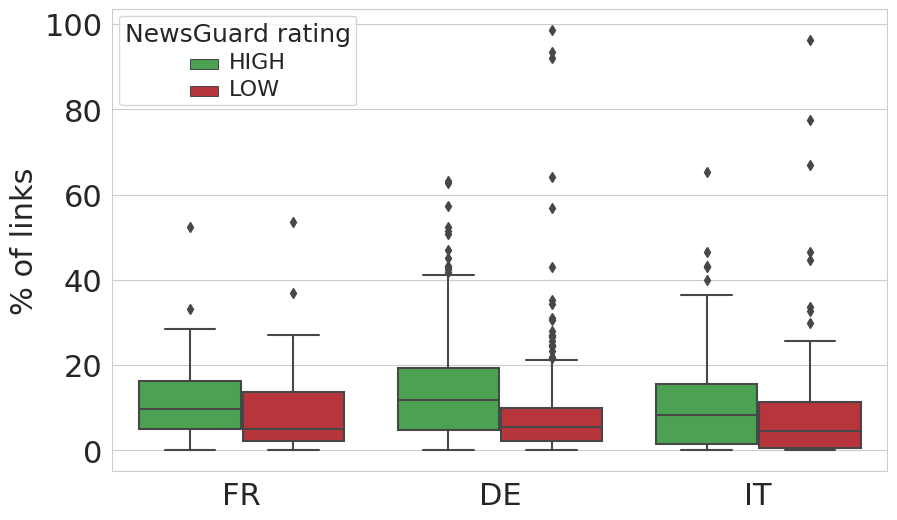}
     \caption{Distributions of the percentage of links per Telegram channel linking to high and low credibility news websites for each dataset. Median HIGH NewsGuard rating \emph{FR} 9.7, \emph{DE} 11.8 and \emph{IT} 8.4, median LOW NewsGuard rating \emph{FR} 5.1, \emph{DE} 5.5 and \emph{IT} 4.6}\label{fig:telegram_boxplot}
\end{figure}

Figure \ref{fig:telegram_boxplot} shows that links sharing news articles within Telegram channels tend to outnumber unreliable ones. The distance between the medians is, however, smaller than the data observed in Figure \ref{fig:newsguard_boxplot}.

\section{Discussion}
\subsection{Conclusions and future works}
We carried out a detailed study of Twitter conversations in French, German, and Italian related to the COVID-19 vaccination campaigns between 2020 and 2021.
We investigated the trustworthiness of news domains shared in this period, identifying a low but non-negligible prevalence of tweets sharing unreliable sources compared to mainstream and reputable sources (which are shared 2-4x more times). We also studied the formation of pro-vaxxer and anti-vaxxer user groups with their respective echo-chambers over time.

The findings from RQ1 indicate that the German language dataset exhibits a distinct pattern compared to other datasets, with a notable absence of significant polarization and a more homogeneous discussion regarding COVID-19, which polarizes only during the end of the period of analysis.
To confirm this hypothesis, we have shown how there are substantial differences in the polarization of users in the different datasets, shown again how the German language dataset tends to polarize after a time interval compared to the other datasets.
However, we can take note that although there are differences in the discussion given by political and behavioural aspects, we have shown how all datasets tend to converge into a well-defined structure despite completely different starting points and communities.\\
The construction of retweet networks allows us to generalize the dynamics of interaction and information dissemination on platforms such as Twitter. This approach allows us to analyze patterns of connection between users, trends in content dissemination, and the evolution of conversations over time.\\
Analysis of retweet networks reveals how, despite the different cultural origins and specific issues addressed in the various datasets, entirely similar structures emerge at the last snapshot. These converging patterns suggest that there are universal mechanisms that drive the dissemination of information and the formation of communities within Twitter, although discussions may differ depending on context.\\
Through the analysis of retweet networks, we are able to gain a comprehensive and deep view of the social and informational dynamics on Twitter, transcending language and cultural barriers to identify the trends and behaviors that characterize this global communication platform on a given topic.\\
Regarding the prevalence of low-credibility information and how they are circulated (RQ2), we explored the credibility of the links to external websites shared in the tweets.
By using NewsGuard to score the credibility of the linked domains, we observed again an evolution: The percentage of low-credibility domains shared within the anti-vaxxer communities is already dominant, like in the French dataset, or increases with time to prevail, like in the Italian and German discussion. For the pro-vaxxer communities, we found an high prevalence of  high-credibility links, stable in the French and Italian datasets or with a tendency to improve in the German discussion. Not surprisingly, we also observed a growth of the  links to external social
media, like YouTube and Telegram that are a clear indication of increasing disinformation, as a symptom of the rise of polarization. \\
In the future, we also aim to extend our comprehensive analyses by considering extending the temporal scope of the study to more recent data would provide a deeper understanding of the evolution of vaccine discourse over time. 
It would also be useful in addition to the three major EU languages, examining vaccine discourse in other languages and regions could highlight broader cross-cultural patterns. This could help identify patterns that we can call global and generalize user behaviors in a given context such as vaccines.

\subsection{Limitations}
Our work presents some limitations. First, the set of keywords used for data collection could introduce a degree of bias, favoring certain perspectives over others related to the vaccines. Additionally, our choice to obtain the data using only keywords might result in an incomplete representation of the topic, omitting crucial nuances and emerging trends that might not be covered by the selected terms as well as linguistic and cultural contexts that employ different terminologies.\\
We rely on data collected through Twitter's Historical Search API V2 for all three datasets in order to ensure overall consistency in the collected data. However, it is important to note that this choice involves a trade-off. While it provides us with stability and consistency in the data over time, it results in the exclusion of suspended users or removed content that would have been obtained through the use of the Filter API in real-time streaming mode. 
As a result, the possible absence of such removed or suspended content emerges as a limitation of our investigation.\\
An additional limitation of our study arises from the use of keywords in three specific languages: Italian, French, and German. Although this has enabled us to gather data across a wide range of countries and regions where these languages are spoken, it is essential to recognize that this methodology might capture not only content from the native countries of these languages but also from nations where they are used as secondary or intercultural communication languages.
A clear example is provided by the dataset in the French language, which exhibits a significantly higher number of activities compared to the others. This data is undoubtedly influenced by the presence of numerous French-speaking countries worldwide.
As a result, the content collected could reflect both the realities of the languages' countries of origin and those of the regions where they are employed.

\subsection{Ethical Considerations}
We did not attempt to identify or deanonymize individual users, and we do not share any personal information and political leanings reporting only public content and figures.
The data presented was collected through public APIs in an aggregate manner, and the ID of tweets is provided transparently so that access can be granted in full accordance with the platforms' data-sharing policies; except for posts that have been removed or made private by users, which hinders reproducibility analyses.
At the time of this writing, we acknowledge that Twitter has limited access to its public API, and we encourage interested researchers to contact us should they be interested in working with our datasets.


\bibliography{main}


\begin{thebibliography}{46}


\ifx \showCODEN    \undefined \def \showCODEN     #1{\unskip}     \fi
\ifx \showDOI      \undefined \def \showDOI       #1{#1}\fi
\ifx \showISBNx    \undefined \def \showISBNx     #1{\unskip}     \fi
\ifx \showISBNxiii \undefined \def \showISBNxiii  #1{\unskip}     \fi
\ifx \showISSN     \undefined \def \showISSN      #1{\unskip}     \fi
\ifx \showLCCN     \undefined \def \showLCCN      #1{\unskip}     \fi
\ifx \shownote     \undefined \def \shownote      #1{#1}          \fi
\ifx \showarticletitle \undefined \def \showarticletitle #1{#1}   \fi
\ifx \showURL      \undefined \def \showURL       {\relax}        \fi
\providecommand\bibfield[2]{#2}
\providecommand\bibinfo[2]{#2}
\providecommand\natexlab[1]{#1}
\providecommand\showeprint[2][]{arXiv:#2}

\bibitem[Aliaksandr~Herasimenka and Howard(2023)]%
        {herasimenka2023telegram}
\bibfield{author}{\bibinfo{person}{Aleksi~Knuutila Aliaksandr~Herasimenka, Jonathan~Bright} {and} \bibinfo{person}{Philip~N. Howard}.} \bibinfo{year}{2023}\natexlab{}.
\newblock \showarticletitle{Misinformation and professional news on largely unmoderated platforms: the case of telegram}.
\newblock \bibinfo{journal}{\emph{Journal of Information Technology \& Politics}} \bibinfo{volume}{20}, \bibinfo{number}{2} (\bibinfo{year}{2023}), \bibinfo{pages}{198--212}.
\newblock
\urldef\tempurl%
\url{https://doi.org/10.1080/19331681.2022.2076272}
\showDOI{\tempurl}


\bibitem[Andrews(2019)]%
        {newslikevirus}
\bibfield{author}{\bibinfo{person}{Edmund Andrews}.} \bibinfo{year}{2019}\natexlab{}.
\newblock \bibinfo{title}{How fake news spreads like a real virus}.
\newblock
\newblock
\urldef\tempurl%
\url{https://engineering.stanford.edu/magazine/article/how-fake-news-spreads-real-virus}
\showURL{%
\tempurl}


\bibitem[Blondel et~al\mbox{.}(2008)]%
        {blondel2008fast}
\bibfield{author}{\bibinfo{person}{Vincent~D Blondel}, \bibinfo{person}{Jean-Loup Guillaume}, \bibinfo{person}{Renaud Lambiotte}, {and} \bibinfo{person}{Etienne Lefebvre}.} \bibinfo{year}{2008}\natexlab{}.
\newblock \showarticletitle{Fast unfolding of communities in large networks}.
\newblock \bibinfo{journal}{\emph{Journal of Statistical Mechanics: Theory and Experiment}} \bibinfo{volume}{2008}, \bibinfo{number}{10} (\bibinfo{year}{2008}).
\newblock


\bibitem[Cinelli et~al\mbox{.}(2020)]%
        {cinelli2020covid}
\bibfield{author}{\bibinfo{person}{Matteo Cinelli}, \bibinfo{person}{Walter Quattrociocchi}, \bibinfo{person}{Alessandro Galeazzi}, \bibinfo{person}{Carlo~Michele Valensise}, \bibinfo{person}{Emanuele Brugnoli}, \bibinfo{person}{Ana~Lucia Schmidt}, \bibinfo{person}{Paola Zola}, \bibinfo{person}{Fabiana Zollo}, {and} \bibinfo{person}{Antonio Scala}.} \bibinfo{year}{2020}\natexlab{}.
\newblock \showarticletitle{The COVID-19 social media infodemic}.
\newblock \bibinfo{journal}{\emph{Scientific reports}} \bibinfo{volume}{10}, \bibinfo{number}{1} (\bibinfo{year}{2020}), \bibinfo{pages}{1--10}.
\newblock
\urldef\tempurl%
\url{https://doi.org/10.1038/s41598-020-73510-5}
\showDOI{\tempurl}


\bibitem[Cossard et~al\mbox{.}(2020)]%
        {cossard2020echo}
\bibfield{author}{\bibinfo{person}{Alessandro Cossard}, \bibinfo{person}{Gianmarco De~Francisci~Morales}, \bibinfo{person}{Kyriaki Kalimeri}, \bibinfo{person}{Yelena Mejova}, \bibinfo{person}{Daniela Paolotti}, {and} \bibinfo{person}{Michele Starnini}.} \bibinfo{year}{2020}\natexlab{}.
\newblock \showarticletitle{Falling into the Echo Chamber: The Italian Vaccination Debate on Twitter}.
\newblock \bibinfo{journal}{\emph{Proceedings of the International AAAI Conference on Web and Social Media}} \bibinfo{volume}{14}, \bibinfo{number}{1} (\bibinfo{date}{May} \bibinfo{year}{2020}), \bibinfo{pages}{130--140}.
\newblock
\urldef\tempurl%
\url{https://doi.org/10.1609/icwsm.v14i1.7285}
\showDOI{\tempurl}


\bibitem[Crupi et~al\mbox{.}(2022)]%
        {crupi2022echoes}
\bibfield{author}{\bibinfo{person}{Giuseppe Crupi}, \bibinfo{person}{Yelena Mejova}, \bibinfo{person}{Michele Tizzani}, \bibinfo{person}{Daniela Paolotti}, {and} \bibinfo{person}{André Panisson}.} \bibinfo{year}{2022}\natexlab{}.
\newblock \showarticletitle{Echoes through Time: Evolution of the Italian COVID-19 Vaccination Debate}.
\newblock \bibinfo{journal}{\emph{Proceedings of the International AAAI Conference on Web and Social Media}} \bibinfo{volume}{16}, \bibinfo{number}{1} (\bibinfo{date}{May} \bibinfo{year}{2022}), \bibinfo{pages}{102--113}.
\newblock
\urldef\tempurl%
\url{https://doi.org/10.1609/icwsm.v16i1.19276}
\showDOI{\tempurl}


\bibitem[Curley et~al\mbox{.}(2022)]%
        {cliona2022telegram}
\bibfield{author}{\bibinfo{person}{Cliona Curley}, \bibinfo{person}{Eugenia Siapera}, {and} \bibinfo{person}{Joe Carthy}.} \bibinfo{year}{2022}\natexlab{}.
\newblock \showarticletitle{Covid-19 Protesters and the Far Right on Telegram: Co-Conspirators or Accidental Bedfellows?}
\newblock \bibinfo{journal}{\emph{Social Media + Society}} \bibinfo{volume}{8}, \bibinfo{number}{4} (\bibinfo{year}{2022}).
\newblock
\urldef\tempurl%
\url{https://doi.org/10.1177/20563051221129187}
\showDOI{\tempurl}


\bibitem[Dias et~al\mbox{.}(2020)]%
        {dias2020social}
\bibfield{author}{\bibinfo{person}{Fernanda~M Dias}, \bibinfo{person}{Bruno~P Rodrigues}, \bibinfo{person}{Luiz~G Araújo}, {and} \bibinfo{person}{Ana~MS Soares}.} \bibinfo{year}{2020}\natexlab{}.
\newblock \showarticletitle{Social media influence in the COVID-19 pandemic}.
\newblock \bibinfo{journal}{\emph{Brazilian Journal of Urology}} \bibinfo{volume}{46}, \bibinfo{number}{6} (\bibinfo{year}{2020}), \bibinfo{pages}{1015--1022}.
\newblock
\urldef\tempurl%
\url{https://doi.org/10.1590/S1677-56922020000600002}
\showDOI{\tempurl}


\bibitem[Donzelli et~al\mbox{.}(2018)]%
        {donzelli2018vaccine}
\bibfield{author}{\bibinfo{person}{Gabriele Donzelli}, \bibinfo{person}{Giacomo Palomba}, \bibinfo{person}{Ileana Federigi}, \bibinfo{person}{Francesco Aquino}, \bibinfo{person}{Lorenzo Cioni}, \bibinfo{person}{Marco Verani}, \bibinfo{person}{Annalaura Carducci}, {and} \bibinfo{person}{Pierluigi Lopalco}.} \bibinfo{year}{2018}\natexlab{}.
\newblock \showarticletitle{Misinformation on vaccination: A quantitative analysis of YouTube videos}.
\newblock \bibinfo{journal}{\emph{Human Vaccines \& Immunotherapeutics}} \bibinfo{volume}{14}, \bibinfo{number}{7} (\bibinfo{year}{2018}), \bibinfo{pages}{1654--1659}.
\newblock
\urldef\tempurl%
\url{https://doi.org/10.1080/21645515.2018.1454572}
\showDOI{\tempurl}


\bibitem[Elias(2021)]%
        {elias2021youtube}
\bibfield{author}{\bibinfo{person}{John Elias}.} \bibinfo{year}{2021}\natexlab{}.
\newblock \bibinfo{title}{YouTube to Add Labels to Some Health Videos amid Misinformation Backlash}.
\newblock \bibinfo{howpublished}{\url{https://www.cnbc.com/2021/07/19/youtube-labeling-some-health-videos-amid-misinformation-backlash.html}}.
\newblock


\bibitem[for Broadband \&~Society(2017)]%
        {SocialFreespeechMisinfo}
\bibfield{author}{\bibinfo{person}{Benton~Institute for Broadband \&~Society}.} \bibinfo{year}{2017}\natexlab{}.
\newblock \bibinfo{booktitle}{\emph{The Future of Free Speech, Trolls, Anonymity, and Fake News Online}}.
\newblock
\urldef\tempurl%
\url{https://www.benton.org/headlines/future-free-speech-trolls-anonymity-and-fake-news-online}
\showURL{%
\tempurl}


\bibitem[Gatta et~al\mbox{.}(2023)]%
        {gatta2023interconnected}
\bibfield{author}{\bibinfo{person}{Valerio~La Gatta}, \bibinfo{person}{Luca Luceri}, \bibinfo{person}{Francesco Fabbri}, {and} \bibinfo{person}{Emilio Ferrara}.} \bibinfo{year}{2023}\natexlab{}.
\newblock \showarticletitle{The interconnected nature of online harm and moderation: investigating the cross-platform spread of harmful content between youtube and Twitter}. In \bibinfo{booktitle}{\emph{Proceedings of the 34th ACM conference on hypertext and social media}}. \bibinfo{pages}{1--10}.
\newblock


\bibitem[Germani and Biller-Andorno(2021)]%
        {germani2021anti}
\bibfield{author}{\bibinfo{person}{Federico Germani} {and} \bibinfo{person}{Nikola Biller-Andorno}.} \bibinfo{year}{2021}\natexlab{}.
\newblock \showarticletitle{The anti-vaccination infodemic on social media: A behavioral analysis}.
\newblock \bibinfo{journal}{\emph{PLoS One}} \bibinfo{volume}{16}, \bibinfo{number}{3} (\bibinfo{year}{2021}).
\newblock
\urldef\tempurl%
\url{https://doi.org/10.1371/journal.pone.0247642}
\showDOI{\tempurl}


\bibitem[Ginossar et~al\mbox{.}(2022)]%
        {ginossar2022cross}
\bibfield{author}{\bibinfo{person}{Tamar Ginossar}, \bibinfo{person}{Iain~J Cruickshank}, \bibinfo{person}{Elena Zheleva}, \bibinfo{person}{Jason Sulskis}, {and} \bibinfo{person}{Tanya Berger-Wolf}.} \bibinfo{year}{2022}\natexlab{}.
\newblock \showarticletitle{Cross-platform spread: Vaccine-related content, sources, and conspiracy theories in {YouTube} videos shared in early {Twitter COVID-19} conversations}.
\newblock \bibinfo{journal}{\emph{Human Vaccines \& Immunotherapeutics}} \bibinfo{volume}{18}, \bibinfo{number}{1} (\bibinfo{year}{2022}), \bibinfo{pages}{1--13}.
\newblock
\urldef\tempurl%
\url{https://doi.org/10.1080/21645515.2021.2003647}
\showDOI{\tempurl}


\bibitem[Giovanni et~al\mbox{.}(2022)]%
        {giovanni2022vaccineu}
\bibfield{author}{\bibinfo{person}{Marco~Di Giovanni}, \bibinfo{person}{Francesco Pierri}, \bibinfo{person}{Christopher Torres-Lugo}, {and} \bibinfo{person}{Marco Brambilla}.} \bibinfo{year}{2022}\natexlab{}.
\newblock \showarticletitle{VaccinEU: COVID-19 Vaccine Conversations on Twitter in French, German and Italian}.
\newblock \bibinfo{journal}{\emph{Proceedings of the International AAAI Conference on Web and Social Media}} (\bibinfo{year}{2022}).
\newblock
\urldef\tempurl%
\url{https://doi.org/10.1609/icwsm.v16i1.19374}
\showDOI{\tempurl}


\bibitem[Guntuku et~al\mbox{.}(2021)]%
        {guntuku2021temporal}
\bibfield{author}{\bibinfo{person}{Sharath~Chandra Guntuku}, \bibinfo{person}{Alison~M Buttenheim}, \bibinfo{person}{Garrick Sherman}, {and} \bibinfo{person}{Raina~M Merchant}.} \bibinfo{year}{2021}\natexlab{}.
\newblock \showarticletitle{Twitter discourse reveals geographical and temporal variation in concerns about COVID-19 vaccines in the United States}.
\newblock \bibinfo{journal}{\emph{Vaccine}} \bibinfo{volume}{39}, \bibinfo{number}{30} (\bibinfo{year}{2021}), \bibinfo{pages}{4034--4038}.
\newblock


\bibitem[Hussein et~al\mbox{.}(2020)]%
        {hussein2020measuring}
\bibfield{author}{\bibinfo{person}{Eslam Hussein}, \bibinfo{person}{Prerna Juneja}, {and} \bibinfo{person}{Tanushree Mitra}.} \bibinfo{year}{2020}\natexlab{}.
\newblock \showarticletitle{Measuring misinformation in video search platforms: An audit study on {YouTube}}.
\newblock \bibinfo{journal}{\emph{Proceedings of the ACM on Human-Computer Interaction}} \bibinfo{volume}{4}, \bibinfo{number}{CSCW1} (\bibinfo{year}{2020}), \bibinfo{pages}{1--27}.
\newblock


\bibitem[Jemielniak and Krempovych(2021)]%
        {JEMIELNIAK20214}
\bibfield{author}{\bibinfo{person}{D. Jemielniak} {and} \bibinfo{person}{Y. Krempovych}.} \bibinfo{year}{2021}\natexlab{}.
\newblock \showarticletitle{An analysis of AstraZeneca COVID-19 vaccine misinformation and fear mongering on Twitter}.
\newblock \bibinfo{journal}{\emph{Public Health}}  \bibinfo{volume}{200} (\bibinfo{year}{2021}), \bibinfo{pages}{4--6}.
\newblock
\showISSN{0033-3506}
\urldef\tempurl%
\url{https://doi.org/10.1016/j.puhe.2021.08.019}
\showDOI{\tempurl}


\bibitem[Jiang et~al\mbox{.}(2020)]%
        {jiang2020political}
\bibfield{author}{\bibinfo{person}{Julie Jiang}, \bibinfo{person}{Emily Chen}, \bibinfo{person}{Shen Yan}, \bibinfo{person}{Kristina Lerman}, {and} \bibinfo{person}{Emilio Ferrara}.} \bibinfo{year}{2020}\natexlab{}.
\newblock \showarticletitle{Political polarization drives online conversations about COVID-19 in the United States}.
\newblock \bibinfo{journal}{\emph{Human Behavior and Emerging Technologies}} \bibinfo{volume}{2}, \bibinfo{number}{3} (\bibinfo{year}{2020}), \bibinfo{pages}{200--211}.
\newblock
\urldef\tempurl%
\url{https://doi.org/10.1002/hbe2.202}
\showDOI{\tempurl}


\bibitem[Jiang et~al\mbox{.}(2021)]%
        {jiang2021social}
\bibfield{author}{\bibinfo{person}{Julie Jiang}, \bibinfo{person}{Xiang Ren}, \bibinfo{person}{Emilio Ferrara}, {et~al\mbox{.}}} \bibinfo{year}{2021}\natexlab{}.
\newblock \showarticletitle{Social media polarization and echo chambers in the context of {COVID-19}: Case study}.
\newblock \bibinfo{journal}{\emph{JMIRx Med}} \bibinfo{volume}{2}, \bibinfo{number}{3} (\bibinfo{year}{2021}).
\newblock


\bibitem[Kiran et~al\mbox{.}(2018)]%
        {garimella2018randomwalk}
\bibfield{author}{\bibinfo{person}{Garimella Kiran}, \bibinfo{person}{De~Francisci~Morales Gianmarco}, \bibinfo{person}{Gionis Aristides}, {and} \bibinfo{person}{Mathioudakis Michael}.} \bibinfo{year}{2018}\natexlab{}.
\newblock \showarticletitle{Quantifying Controversy on Social Media}.
\newblock \bibinfo{journal}{\emph{ACM Transactions on Social Computing}} (\bibinfo{year}{2018}).
\newblock


\bibitem[Knuutila et~al\mbox{.}(2020)]%
        {knuutila2020covid}
\bibfield{author}{\bibinfo{person}{Aleksi Knuutila}, \bibinfo{person}{Aliaksandr Herasimenka}, \bibinfo{person}{Hubert Au}, \bibinfo{person}{Jonathan Bright}, \bibinfo{person}{Rasmus Nielsen}, \bibinfo{person}{Philip~N Howard}, {et~al\mbox{.}}} \bibinfo{year}{2020}\natexlab{}.
\newblock \showarticletitle{{COVID-related misinformation on YouTube: The spread of misinformation videos on social media and the effectiveness of platform policies}}.
\newblock \bibinfo{journal}{\emph{COMPROP Data Memo 2020.6}} (\bibinfo{year}{2020}).
\newblock


\bibitem[Krishnan et~al\mbox{.}(2021)]%
        {krishnan2021research}
\bibfield{author}{\bibinfo{person}{Nandita Krishnan}, \bibinfo{person}{Jiayan Gu}, \bibinfo{person}{Rebekah Tromble}, {and} \bibinfo{person}{Lorien~C Abroms}.} \bibinfo{year}{2021}\natexlab{}.
\newblock \showarticletitle{Research note: Examining how various social media platforms have responded to {COVID-19} misinformation}.
\newblock \bibinfo{journal}{\emph{Harvard Kennedy School (HKS) Misinformation Review}} \bibinfo{volume}{2}, \bibinfo{number}{6} (\bibinfo{year}{2021}), \bibinfo{pages}{1--25}.
\newblock


\bibitem[Lenti et~al\mbox{.}(2023)]%
        {lenti2023misinformation}
\bibfield{author}{\bibinfo{person}{Jacopo Lenti}, \bibinfo{person}{Yelena Mejova}, \bibinfo{person}{Kyriaki Kalimeri}, \bibinfo{person}{Andr{\'e} Panisson}, \bibinfo{person}{Daniela Paolotti}, \bibinfo{person}{Michele Tizzani}, {and} \bibinfo{person}{Michele Starnini}.} \bibinfo{year}{2023}\natexlab{}.
\newblock \showarticletitle{Global Misinformation Spillovers in the Vaccination Debate Before and During the COVID-19 Pandemic: Multilingual Twitter Study}.
\newblock \bibinfo{journal}{\emph{JMIR Infodemiology}}  \bibinfo{volume}{3} (\bibinfo{date}{24 May} \bibinfo{year}{2023}), \bibinfo{pages}{e44714}.
\newblock
\showISSN{2564-1891}
\urldef\tempurl%
\url{https://doi.org/10.2196/44714}
\showDOI{\tempurl}


\bibitem[Li et~al\mbox{.}(2020)]%
        {lie2020youtube}
\bibfield{author}{\bibinfo{person}{Heidi Oi-Yee Li}, \bibinfo{person}{Adrian Bailey}, \bibinfo{person}{David Huynh}, {and} \bibinfo{person}{James Chan}.} \bibinfo{year}{2020}\natexlab{}.
\newblock \showarticletitle{YouTube as a source of information on {COVID-19}: A pandemic of misinformation?}
\newblock \bibinfo{journal}{\emph{BMJ Global Health}} \bibinfo{volume}{5}, \bibinfo{number}{5} (\bibinfo{year}{2020}).
\newblock


\bibitem[Loomba et~al\mbox{.}(2021)]%
        {loomba2021measuring}
\bibfield{author}{\bibinfo{person}{Sahil Loomba}, \bibinfo{person}{Alexandre de Figueiredo}, \bibinfo{person}{Simon~J Piatek}, \bibinfo{person}{Kristen de Graaf}, {and} \bibinfo{person}{Heidi~J Larson}.} \bibinfo{year}{2021}\natexlab{}.
\newblock \showarticletitle{Measuring the impact of {COVID-19} vaccine misinformation on vaccination intent in the {UK} and {USA}}.
\newblock \bibinfo{journal}{\emph{Nature Human Behaviour}} \bibinfo{volume}{5}, \bibinfo{number}{3} (\bibinfo{year}{2021}), \bibinfo{pages}{337--348}.
\newblock


\bibitem[Luceri et~al\mbox{.}(2021)]%
        {luceri2021social}
\bibfield{author}{\bibinfo{person}{Luca Luceri}, \bibinfo{person}{Stefano Cresci}, {and} \bibinfo{person}{Silvia Giordano}.} \bibinfo{year}{2021}\natexlab{}.
\newblock \showarticletitle{{Social media against society: Information manipulation in the 2020 election}}.
\newblock In \bibinfo{booktitle}{\emph{The Internet and the 2020 election}}, \bibfield{editor}{\bibinfo{person}{Jody Baumgartner} {and} \bibinfo{person}{Terri Towner}} (Eds.). \bibinfo{pages}{3--23}.
\newblock


\bibitem[Malova(2021)]%
        {malova2021vaccine}
\bibfield{author}{\bibinfo{person}{Ekaterina Malova}.} \bibinfo{year}{2021}\natexlab{}.
\newblock \showarticletitle{Understanding online conversations about COVID-19 vaccine on Twitter: vaccine hesitancy amid the public health crisis}.
\newblock \bibinfo{journal}{\emph{Communication Research Reports}} \bibinfo{volume}{38}, \bibinfo{number}{5} (\bibinfo{year}{2021}), \bibinfo{pages}{346--356}.
\newblock


\bibitem[Memon and Carley(2020)]%
        {memon2020}
\bibfield{author}{\bibinfo{person}{Shahan~Ali Memon} {and} \bibinfo{person}{Kathleen~M. Carley}.} \bibinfo{year}{2020}\natexlab{}.
\newblock \showarticletitle{Characterizing COVID-19 Misinformation Communities Using a Novel Twitter Dataset}.
\newblock \bibinfo{journal}{\emph{arXiv preprint arXiv:2008.00791}} (\bibinfo{year}{2020}).
\newblock
\urldef\tempurl%
\url{https://doi.org/10.48550/arXiv.2008.00791}
\showDOI{\tempurl}


\bibitem[Metaxas et~al\mbox{.}(2015)]%
        {retweetsindicate}
\bibfield{author}{\bibinfo{person}{Panagiotis Metaxas}, \bibinfo{person}{Eni Mustafaraj}, \bibinfo{person}{Kily Wong}, \bibinfo{person}{Laura Zeng}, \bibinfo{person}{Megan O'Keefe}, {and} \bibinfo{person}{Samantha Finn}.} \bibinfo{year}{2015}\natexlab{}.
\newblock \showarticletitle{What Do Retweets Indicate? Results from User Survey and Meta-Review of Research}.
\newblock \bibinfo{journal}{\emph{Proceedings of the International AAAI Conference on Web and Social Media}}  \bibinfo{volume}{9} (\bibinfo{year}{2015}), \bibinfo{pages}{658–661}.
\newblock
\urldef\tempurl%
\url{https://ojs.aaai.org/index.php/ICWSM/article/view/14661#:~:text=Our%20findings%20indicate%20that%20retweeting}
\showURL{%
\tempurl}


\bibitem[Metaxas et~al\mbox{.}(2014)]%
        {metaxas2014retweets}
\bibfield{author}{\bibinfo{person}{Panagiotis~Takis Metaxas}, \bibinfo{person}{Eni Mustafaraj}, \bibinfo{person}{Kily Wong}, \bibinfo{person}{Laura Zeng}, \bibinfo{person}{Megan O'Keefe}, {and} \bibinfo{person}{Samantha Finn}.} \bibinfo{year}{2014}\natexlab{}.
\newblock \bibinfo{title}{Do Retweets indicate Interest, Trust, Agreement? (Extended Abstract)}.
\newblock
\newblock
\showeprint[arxiv]{1411.3555}~[cs.SI]


\bibitem[NewsGuard(2023)]%
        {newsguardprocesscriteria}
\bibfield{author}{\bibinfo{person}{NewsGuard}.} \bibinfo{year}{2023}\natexlab{}.
\newblock \bibinfo{booktitle}{\emph{Rating Process \& Criteria}}.
\newblock NewsGuard.
\newblock
\urldef\tempurl%
\url{https://www.newsguardtech.com/ratings/rating-process-criteria/}
\showURL{%
\tempurl}


\bibitem[Ng and Loke(2020)]%
        {ng2020analyzing}
\bibfield{author}{\bibinfo{person}{Lynnette Hui~Xian Ng} {and} \bibinfo{person}{Jia~Yuan Loke}.} \bibinfo{year}{2020}\natexlab{}.
\newblock \showarticletitle{Analyzing public opinion and misinformation in a {COVID-19} {Telegram} group chat}.
\newblock \bibinfo{journal}{\emph{IEEE Internet Computing}} \bibinfo{volume}{25}, \bibinfo{number}{2} (\bibinfo{year}{2020}), \bibinfo{pages}{84--91}.
\newblock


\bibitem[Nogara et~al\mbox{.}(2023)]%
        {nogara2023toxic}
\bibfield{author}{\bibinfo{person}{Gianluca Nogara}, \bibinfo{person}{Francesco Pierri}, \bibinfo{person}{Stefano Cresci}, \bibinfo{person}{Luca Luceri}, \bibinfo{person}{Petter T{\"o}rnberg}, {and} \bibinfo{person}{Silvia Giordano}.} \bibinfo{year}{2023}\natexlab{}.
\newblock \showarticletitle{Toxic Bias: Perspective API misreads German as more toxic}.
\newblock \bibinfo{journal}{\emph{arXiv preprint arXiv:2312.12651}} (\bibinfo{year}{2023}).
\newblock


\bibitem[Nogara et~al\mbox{.}(2022)]%
        {nogara2022disinformation}
\bibfield{author}{\bibinfo{person}{Gianluca Nogara}, \bibinfo{person}{Padinjaredath~Suresh Vishnuprasad}, \bibinfo{person}{Felipe Cardoso}, \bibinfo{person}{Omran Ayoub}, \bibinfo{person}{Silvia Giordano}, {and} \bibinfo{person}{Luca Luceri}.} \bibinfo{year}{2022}\natexlab{}.
\newblock \showarticletitle{The Disinformation Dozen: An Exploratory Analysis of Covid-19 Disinformation Proliferation on Twitter}. In \bibinfo{booktitle}{\emph{14th ACM Web Science Conference 2022}}. \bibinfo{pages}{348--358}.
\newblock


\bibitem[Pierri(2020)]%
        {pierri2020diffusion}
\bibfield{author}{\bibinfo{person}{Francesco Pierri}.} \bibinfo{year}{2020}\natexlab{}.
\newblock \showarticletitle{The diffusion of mainstream and disinformation news on Twitter: the case of Italy and France}. In \bibinfo{booktitle}{\emph{Companion proceedings of the web conference 2020}}. \bibinfo{pages}{617--622}.
\newblock


\bibitem[Pierri et~al\mbox{.}(2023)]%
        {pierri2023one}
\bibfield{author}{\bibinfo{person}{Francesco Pierri}, \bibinfo{person}{Matthew~R DeVerna}, \bibinfo{person}{Kai-Cheng Yang}, \bibinfo{person}{David Axelrod}, \bibinfo{person}{John Bryden}, {and} \bibinfo{person}{Filippo Menczer}.} \bibinfo{year}{2023}\natexlab{}.
\newblock \showarticletitle{One Year of COVID-19 Vaccine Misinformation on Twitter: Longitudinal Study}.
\newblock \bibinfo{journal}{\emph{Journal of Medical Internet Research}}  \bibinfo{volume}{25} (\bibinfo{year}{2023}), \bibinfo{pages}{e42227}.
\newblock


\bibitem[Reuters(2019)]%
        {reuters2019mediaguard}
\bibfield{author}{\bibinfo{person}{Reuters}.} \bibinfo{year}{2019}\natexlab{}.
\newblock \showarticletitle{NewsGuard to help advertisers avoid misinformation}.
\newblock \bibinfo{journal}{\emph{Reuters}} (\bibinfo{year}{2019}).
\newblock
\urldef\tempurl%
\url{https://www.reuters.com/article/us-media-newsguard-idUSKCN1PQ5FV}
\showURL{%
\tempurl}


\bibitem[Santiago et~al\mbox{.}(2022)]%
        {santiago2022covid}
\bibfield{author}{\bibinfo{person}{Rafael Santiago}, \bibinfo{person}{Saira Ijaz}, \bibinfo{person}{Mohamed El-Sayed}, {and} \bibinfo{person}{Amir Mian}.} \bibinfo{year}{2022}\natexlab{}.
\newblock \showarticletitle{The COVID-19 vaccine social media infodemic: healthcare providers' missed dose in addressing misinformation and vaccine hesitancy}.
\newblock \bibinfo{journal}{\emph{Human Vaccines \& Immunotherapeutics}} \bibinfo{volume}{18}, \bibinfo{number}{1} (\bibinfo{year}{2022}), \bibinfo{pages}{154--161}.
\newblock
\urldef\tempurl%
\url{https://doi.org/10.1080/21645515.2021.1912551}
\showDOI{\tempurl}


\bibitem[Tardelli et~al\mbox{.}(2023)]%
        {tardelli2023temporal}
\bibfield{author}{\bibinfo{person}{Serena Tardelli}, \bibinfo{person}{Leonardo Nizzoli}, \bibinfo{person}{Maurizio Tesconi}, \bibinfo{person}{Mauro Conti}, \bibinfo{person}{Preslav Nakov}, \bibinfo{person}{Giovanni Da~San~Martino}, {and} \bibinfo{person}{Stefano Cresci}.} \bibinfo{year}{2023}\natexlab{}.
\newblock \showarticletitle{Temporal dynamics of coordinated online behavior: Stability, archetypes, and influence}.
\newblock \bibinfo{journal}{\emph{arXiv preprint:2301.06774}} (\bibinfo{year}{2023}).
\newblock


\bibitem[Union(2023)]%
        {EUcommonworkCovid}
\bibfield{author}{\bibinfo{person}{European Union}.} \bibinfo{year}{2023}\natexlab{}.
\newblock \bibinfo{booktitle}{\emph{The common EU response to COVID-19}}.
\newblock European Union.
\newblock
\urldef\tempurl%
\url{https://european-union.europa.eu/priorities-and-actions/common-eu-response-covid-19_en}
\showURL{%
\tempurl}


\bibitem[Villa et~al\mbox{.}(2021)]%
        {villa2021echo}
\bibfield{author}{\bibinfo{person}{Giacomo Villa}, \bibinfo{person}{Gabriella Pasi}, {and} \bibinfo{person}{Marco Viviani}.} \bibinfo{year}{2021}\natexlab{}.
\newblock \showarticletitle{Echo chamber detection and analysis: A topology- and content-based approach in the {COVID-19} scenario}.
\newblock \bibinfo{journal}{\emph{Social Network Analysis and Mining}} \bibinfo{volume}{11}, \bibinfo{number}{1} (\bibinfo{year}{2021}), \bibinfo{pages}{78}.
\newblock


\bibitem[Vishnuprasad et~al\mbox{.}(2024)]%
        {vishnuprasad2024tracking}
\bibfield{author}{\bibinfo{person}{Padinjaredath~Suresh Vishnuprasad}, \bibinfo{person}{Gianluca Nogara}, \bibinfo{person}{Felipe Cardoso}, \bibinfo{person}{Silvia Giordano}, \bibinfo{person}{Stefano Cresci}, {and} \bibinfo{person}{Luca Luceri}.} \bibinfo{year}{2024}\natexlab{}.
\newblock \showarticletitle{{Tracking fringe and coordinated activity on Twitter leading up to the US Capitol attack}}. In \bibinfo{booktitle}{\emph{The 18th International AAAI Conference on Web and Social Media (ICWSM'24)}}. AAAI.
\newblock


\bibitem[Wang et~al\mbox{.}(2023)]%
        {wang2023identifying}
\bibfield{author}{\bibinfo{person}{Emily~L Wang}, \bibinfo{person}{Luca Luceri}, \bibinfo{person}{Francesco Pierri}, {and} \bibinfo{person}{Emilio Ferrara}.} \bibinfo{year}{2023}\natexlab{}.
\newblock \showarticletitle{Identifying and characterizing behavioral classes of radicalization within the QAnon conspiracy on Twitter}. In \bibinfo{booktitle}{\emph{Proceedings of the International AAAI Conference on Web and Social Media}}, Vol.~\bibinfo{volume}{17}. \bibinfo{pages}{890--901}.
\newblock


\bibitem[Yang et~al\mbox{.}(2021)]%
        {cheng2021twittervsfacebook}
\bibfield{author}{\bibinfo{person}{Kai-Cheng Yang}, \bibinfo{person}{Francesco Pierri}, \bibinfo{person}{Pik-Mai Hui}, \bibinfo{person}{David Axelrod}, \bibinfo{person}{Christopher Torres-Lugo}, \bibinfo{person}{John Bryden}, {and} \bibinfo{person}{Filippo Menczer}.} \bibinfo{year}{2021}\natexlab{}.
\newblock \showarticletitle{The COVID-19 Infodemic: Twitter versus Facebook}.
\newblock \bibinfo{journal}{\emph{Big Data \& Society}} \bibinfo{volume}{8}, \bibinfo{number}{1} (\bibinfo{year}{2021}), \bibinfo{pages}{20539517211013861}.
\newblock
\urldef\tempurl%
\url{https://doi.org/10.1177/20539517211013861}
\showDOI{\tempurl}


\bibitem[Yuxi et~al\mbox{.}(2019)]%
        {wang2019misinformation}
\bibfield{author}{\bibinfo{person}{Wang Yuxi}, \bibinfo{person}{McKee Martin}, \bibinfo{person}{Torbica Aleksandra}, {and} \bibinfo{person}{Stuckler David}.} \bibinfo{year}{2019}\natexlab{}.
\newblock \showarticletitle{Systematic Literature Review on the Spread of Health-related Misinformation on Social Media}.
\newblock \bibinfo{journal}{\emph{Social Science \& Medicine}} (\bibinfo{year}{2019}).
\newblock


\end{thebibliography}

\end{document}